\documentclass[12pt, draftclsnofoot, onecolumn]{IEEEtran}
\usepackage[noadjust]{cite}

\usepackage[utf8]{inputenc}
\usepackage{amsmath}
\usepackage{amsfonts,amssymb}
\usepackage{array}
\usepackage{graphicx}
\usepackage{subfig}
\usepackage{booktabs}
\usepackage{afterpage}
\usepackage{float}
\usepackage{amsmath}
\usepackage{amsbsy}
\usepackage{mathtools}
\usepackage{color}
\usepackage[thmmarks,amsmath]{ntheorem}
\usepackage[linesnumbered,ruled]{algorithm2e}
\SetKwComment{Comment}{$\triangleright$\ }{}

\DeclarePairedDelimiter{\ceil}{\lceil}{\rceil}

\def\bA{{\mathbf{A}}}  \def\bC{{\mathbf{C}}} \def\bD{{\mathbf{D}}} \def\bE{{\mathbf{E}}}
 \def\bG{{\mathbf{G}}} \def\bH{{\mathbf{H}}}  \def\bJ{{\mathbf{J}}}
    
 \def\bQ{{\mathbf{Q}}}  \def\bS{{\mathbf{S}}} 
\def\bU{{\mathbf{U}}} \def\bV{{\mathbf{V}}} \def\bW{{\mathbf{W}}} \def\bX{{\mathbf{X}}} \def\bY{{\mathbf{Y}}}
\def\bZ{{\mathbf{Z}}}
\def\ba{{\mathbf{a}}}    
\def\bf{{\mathbf{f}}} \def\bg{{\mathbf{g}}}

\def\C{{\mathbb{C}}}

\begin{document}
\title{Two-Stage Channel Estimation for Hybrid RIS Assisted MIMO Systems}
\author{Rafaela~Schroeder,~\IEEEmembership{Graduate Student Member,~IEEE,}
Jiguang~He,~\IEEEmembership{Member,~IEEE,}
Glauber~Brante,~\IEEEmembership{Senior Member,~IEEE,}
and~Markku~Juntti,~\IEEEmembership{Fellow,~IEEE} \\
\thanks{R. Schroeder, J. He and M. Juntti are with the Centre for Wireless Communications- Radio Technologies, FI-90014, University of Oulu, Finland~(e-mail: rafaela.schroeder@oulu.fi, jiguang.he@oulu.fi, markku.juntti@oulu.fi). G. Brante is with the Graduate Program in Electrical and Computer Engineering, Federal University of Technology Paraná, Brazil~(e-mail: gbrante@utfpr.edu.br).}
\thanks{This work has been financially supported in part by the Academy of Finland (ROHM project, grant 319485), European Union's Horizon 2020 Framework Programme for Research and Innovation (ARIADNE project, under grant agreement no. 871464), Academy of Finland 6Genesis Flagship (grant 318927), CAPES (Finance Code 001) and CNPq Brazil.}}
\maketitle

\begin{abstract}
Reconfigurable intelligent surfaces (RISs) have been proposed as a key enabler to improve the coverage of the signals and mitigate the frequent blockages in millimeter wave (mmWave) multiple-input multiple-output (MIMO) communications. However, the channel state information (CSI) acquisition is one of the major challenges for the practical deployment of the RIS. The passive RIS without any baseband processing capabilities brings difficulty on the channel estimation (CE), since the individual channels or the cascaded one can be estimated only at base station (BS) via uplink training or mobile station (MS) via downlink training. In order to facilitate the CSI acquisition, we focus on the hybrid RIS architecture, where a small number of elements are active and able to receive and process the pilot signals at the RIS. The CE is performed in two stages by following the atomic norm minimization to recover the channel parameters, i.e., angles of departure (AoDs), angles of arrival (AoAs), and propagation path gains. Simulation results show that the proposed scheme can outperform the passive RIS CE under the same training overhead. Furthermore, we also study the theoretical performance limits in terms of mean square error (MSE) via Cram\'er-Rao lower bound (CRLB) analyses.    
\end{abstract}


\begin{IEEEkeywords}
Channel estimation, reconfigurable intelligent surface,  Cram\'er-Rao lower bound, mmWave MIMO, hybrid RIS
\end{IEEEkeywords}

\section{Introduction}
\IEEEPARstart{M}{illimeter-wave} (mmWave) multiple-input multiple-output (MIMO) systems are regarded as an essential technology for the fifth generation (5G) wireless networks~\cite{heath2016overview,rappaportmmWave}. In order to compensate for the path loss effect, large antenna arrays are required at both the transmitter and receiver. The features of the channel model in mmWave bands are tightly associated with the high frequencies, which lead to frequent blockage~\cite{raghavan2019statistical}, inevitable high path loss, and inherent channel sparsity. Nevertheless, mmWave MIMO systems heavily rely on line-of-sight (LoS) to  guarantee sufficient received power and spectral efficiency (SE).~\cite{alkhateebCEHybridprecod}

Reconfigurable intelligent surfaces (RISs)~\cite{ozdogan2019intelligent} have been proposed as a promising solution to maintain the coverage of the signals and resolve the blockage problem in mmWave MIMO systems~\cite{zhang2020capacity, he2020,wu2019intelligent,li2020joint,basar2019wireless2,liaskos2018new}. The RIS commonly consists of an array with low-cost discrete phase shifters. By adjusting these phase shifters, the RIS can modify their signal response to achieve a certain objective, for example, focusing the signal towards the receiver~\cite{ardah2020trice,gong2020toward}. The introduction of the RIS can bring other benefits, such as improved physical layer security~\cite{dong2020secure, chu2019secure} and enhanced SE~\cite{zhang2020capacity}. In addition, the RIS can also increase the accuracy of indoor and outdoor localization~\cite{wymeersch2020radio,He2020WCNCW}.  

To enable the optimal or suboptimal joint design of the beamforming vectors at the base station (BS) and mobile station (MS)~\cite{abeywickrama2020intelligent} and RIS phase control matrix, nearly perfect channel state information (CSI) is of great essence for RIS-aided mmWave MIMO systems. The CSI generally includes the individual RIS-BS and MS-RIS channels, cascaded channel, or corresponding channel parameters. However, the CSI acquisition is already a challenging task in mmWave MIMO systems, and even more complicated in RIS-aided scenarios. For instance, in the \textit{passive} RIS without any baseband processing capabilities, the CE can be done only at the BS or MS. Various CE methods for passive RIS have been investigated in~\cite{he2020,ardah2020trice,he2020anm,CE_He_Zhen-Qing,CE_mirza2019,de2021channel,you2020intelligent}. In~\cite{CE_mirza2019}, Mirza and Ali first estimated the direct BS-MS channel, and further applied the bilinear adaptive vector approximate message passing (BAdVAMP) algorithm to obtain the individual RIS-BS and MS-RIS channels. In our recent work~\cite{he2020anm}, we proposed a two-stage CE via atomic norm minimization (ANM) for the passive RIS architecture. At the first CE stage, we aim to estimate the angles of arrival (AoAs) at the MS and angles of departure (AoDs) at the BS. Based on these estimates, we design the beam training matrix and the combing matrix for the second stage sounding. At the second CE stage, we target at recovering the remaining channel parameters, i.e, the angle difference associated with the RIS and products of path gains from the received pilot signals. Note that we keep the beam training matrix and the combining matrix fixed, while we change the RIS phase control matrix during the second stage sounding. With this method, we can reduce the training overhead yet obtain a super-resolution estimation of the channel parameters. 

Alternatively, the \textit{hybrid} RIS with a mix of both \textit{passive} and \textit{active} elements has been proposed by Taha \textit{et al.} in~\cite{ActElements} to reduce the difficulty on CSI acquisition. The active elements are introduced to sense the received signals. With this assumption, the CE can be even performed at the RIS, which simplifies the CE at the sacrifice of  higher power consumption and hardware complexity at the RIS. Motivated by the pioneering work in~\cite{ActElements}, the hybrid architecture has been intensively studied in~\cite{oneRFchain, schroeder2020passive,nguyen2021hybrid,lin2021tensor,alexandropoulos2021hybrid,schroeder2021CEnew}. For instance, Alexandropoulos \textit{et al.} in~\cite{alexandropoulos2021hybrid} described the hardware design of the hybrid RIS and presented a full-wave proof-of-concept of the  application. In our previous work~\cite{schroeder2020passive}, we compared the CE performance between the passive and hybrid RIS architectures for mmWave MIMO systems considering the two-way uplink and downlink training. Interestingly, we noticed that the passive RIS CE can outperform the hybrid RIS CE in the particular scenario of~\cite{schroeder2020passive}, where CE was totally performed at the RIS using a two-way training. Nevertheless, in order to exploit the benefits of the deployment of the hybrid RIS, i.e, mitigation of the CSI acquisition, we keep investigating methods that can further improve CE, outperforming the passive RIS. 

For the reasons discussed above, in this paper we focus on hybrid RIS CE and extend our previous work~\cite{schroeder2021CEnew} in a more comprehensive manner. By adopting the structured channel training, we develop a two-stage CE for hybrid RIS assisted mmWave MIMO systems. Different from our work in~\cite{schroeder2020passive}, we restrict the training to only one way via the uplink transmission. Despite the fact that both works employ the hybrid RIS architecture, the CE formulations are quite distinct. For example, in~\cite{schroeder2020passive}, the CE is performed only at the RIS with a large number of active elements. Here, we develop the training taking in consideration two received signals, one at RIS and the other at BS. We also reduced the number of active elements deployed for CE. 
In the training procedure, the MS sends pilot signals to both RIS and BS. The signals are received at the RIS by the active elements, while the remaining passive elements reflect the signal towards the BS. At the first CE stage, we target at the recovery of the channel parameters in the MS-RIS channel via ANM based on the received signals at the RIS. After that, the estimates of the channel parameters are transmitted to the BS via an error-free backhaul link. Thus, the BS can reconstruct the MS-RIS channel matrix based on the feedback. In order to simplify the CE and reduce the training overhead, we use the reconstructed channel matrix at the second CE stage, where we target at recovering the remaining channel parameters based on the received signals at the BS by resorting to ANM. We evaluate the performance in terms of the mean square error (MSE) of the channel parameter estimation and SE. Our simulation results prove that our proposed CE method not only simplifies the CSI acquisition for RIS-aided mmWave MIMO systems with the aid of a small number of active RIS elements but also brings better CE performance than the passive RIS~\cite{he2020anm} and two-way uplink downlink training~\cite{schroeder2020passive}.  

We would like to emphasize the differences between this paper and our conference paper~\cite{schroeder2021CEnew}. Although we employ the two-stage CE for hybrid RIS for both, in the current work we develop in detail the theoretical limit analysis via CRLB. We clarify that the CRLB illustrates the lower bound for our system model, including the combining matrices, set of active elements and number of training sequences on the estimation performance. In addition, in the current paper we also discuss the location of the RIS and its impact on the CE performance. The contributions of this paper are summarized as follows: 
\begin{itemize}
    \item We propose a two-stage CE procedure for the hybrid RIS-assisted mmWave MIMO systems based on ANM. Since we adopt the hybrid RIS architecture, we can decouple the CE problems as to simplify the CSI acquisition.
    \item We provide a theoretical analysis in terms of the Cramér-Rao lower bound, which serves as the theoretical benchmark for our proposed CE method. 
    \item We study the effect of the RIS location and path loss on the performance of hybrid RIS CE, where the BS and MS locations are fixed while the RIS location varies. The study can be used as a guidance for the RIS deployment in the current and upcoming wireless systems. 
\end{itemize}
The rest of the paper is organized as follows. Section~\ref{sec: system_model} describes the channel model, while Section~\ref{sec:sound_procedure} provides the sounding procedure for the RIS architectures. The proposed CE is detailed in Section~\ref{section: CE}, followed by the design of RIS phase control matrix and BS/MS beamforming vectors in the Section~\ref{sec:beamforming}. Furthermore, the CRLB is studied in the Section~\ref{section: crlb}. The performance evaluation, metrics and simulation results are provided in Section~\ref{section: perfomance} with conclusions drawn in Section~\ref{section: conclusions}.

\textit{Notation}: A bold capital letter $\mathbf{A}$ denotes a matrix and a lowercase letter $\mathbf{a}$ denotes a column vector, and $()^{\mathsf{H}}$, $()^*$, and $()^{\mathsf{T}}$ denote the Hermitian transpose, conjugate, and transpose, respectively. $[\ba]_m$ is the $m$th element of $\ba$, $[\bA]_{:,n}$ and $[\bA]_{m,n}$ are the $n$th column and the $(m,n)$th entry of $\bA$, respectively. $\otimes$ denotes the Kronecker product, $\odot$ is the Khatri-Rao product, $\mathrm{vec} (\mathbf{A})$ is the vectorization of $\mathbf{A}$, $\mathrm{diag}(\mathbf{a})$ being a square diagonal matrix with entries of $\mathbf{a}$ on its diagonal, $\|.\|_{\mathrm{F}}$ is the Frobenius norm, $\mathrm{Tr} (\mathbf{A})$ is the sum value of the diagonal elements of $\mathbf{A}$, $\mathbb{T}(\mathbf{A})$ denotes the block Toeplitz matrix constructed from the vectorized form of $\mathbf{A}$, i.e., $\mathrm{vec}(\mathbf{A})$, being its first row. $(\cdot){\dagger}$ denotes the Moore–Penrose inverse, $\mathcal{A}$ is the atomic set, $\mathrm{conv}({\mathcal{A}})$ denotes the convex hull of $\mathcal{A}$, and $\mathbb{E}$ is the expectation operator. 

\section{Channel Model}
\label{sec: system_model}
We consider the RIS-assisted mmWave MIMO systems, which include one multi-antenna BS, one multi-element RIS, and one multi-antenna MS, as depicted in Fig.~\ref{fig1:RIS}. The number of antennas at the BS and the MS are denoted by $N_\text{B}$ and $N_\text{M}$, respectively, and the number of elements at the RIS is $N_\text{R}$. We assume that the direct BS-MS channel is obstructed (e.g, due to blockage), which leads to the demand on the deployment of the RIS to maintain the connectivity between the BS and MS. We adopt a uniform linear array (ULA) for the antennas/elements, but the extension to an uniform planar array (UPA) is straightforward. Moreover, we focus on a CE proposal requiring only uplink training procedures. 

The propagation channels consist of two tandem channels, i.e., MS-RIS and RIS-BS channels, denoted by $\bH_{\text{M,R}}\in\C^{N_{\text{R}}\times N_{\text{M}}}$ and $\bH_{\text{R,B}}\in\C^{N_{\text{B}}\times N_{\text{R}}}$, respectively. We adopt a block-fading channel, which means that $\bH_{\text{M,R}}$ and $\bH_{\text{R,B}}$ stay constant during a certain period of time, known as coherence time. By following the geometric channel model, we write $\bH_{\text{M,R}}$ as 
\begin{align}
\label{eq:hrm}
    \bH_{\text{M,R}} &= \sum\limits_{l = 1}^{L_{\text{M,R}}} [\boldsymbol{\rho}_{\text{M,R}}]_l \boldsymbol {\alpha}([\boldsymbol{\phi}_{\text{M,R}}]_l ) \boldsymbol{\alpha}^{\mathsf{H}}([\boldsymbol{\theta}_{\text{M,R}}]_l ),\nonumber\\
    &=\bA(\boldsymbol{\phi}_{\text{M,R}})\mathrm{diag}(\boldsymbol{\rho}_{\text{M,R}})\bA^{\mathsf{H}}(\boldsymbol{\theta}_{\text{M,R}}),
\end{align}
where $[\boldsymbol{\rho}_{\text{M,R}}]_l$ is the $l$th propagation path gain, $L_{\text{M,R}}$ is the number of paths, $\boldsymbol{\theta}_{\text{M,R}}$ and $\boldsymbol{\phi}_{\text{M,R}}$, are the AoDs and AoAs of the channel, respectively. Finally, $\boldsymbol{\alpha}([\boldsymbol{\phi}_{\text{M,R}}]_l )$ and $\boldsymbol{\alpha}([\boldsymbol{\theta}_{\text{M,R}}]_l)$ are the array response vectors as a function of $[\boldsymbol{\phi}_{\text{M,R}}]_l$ and $[\boldsymbol{\theta}_{\text{M,R}}]_l$. Considering half-wavelength inter-antenna element spacing, the array response vectors are defined as $[\boldsymbol{ \alpha}([\boldsymbol{\phi}_{\text{M,R}}]_l )]_{n} =  \exp\{
j\pi(n-1)\sin( [\boldsymbol{\phi}_{\text{M,R}}]_l)\}$, for $n=1,\cdots,N_\text{R}$ and $[\boldsymbol{\alpha}([\boldsymbol{\theta}_{\text{M,R}}]_l)]_{n} =  \exp\{j\pi(n-1)\sin( [\boldsymbol{\theta}_{\text{M,R}}]_l)\}$, for $n=1,\cdots,N_\text{M}$, and $j = \sqrt{-1}$. Similarly, the array response matrices $\bA(\boldsymbol{\theta}_{\text{M,R}})$ and $\bA(\boldsymbol{\phi}_{\text{M,R}})$ are defined as 
\begin{equation}
    \bA(\boldsymbol{\theta}_{\text{M,R}}) = \Big[\boldsymbol{\alpha}([\boldsymbol{\theta}_{\text{M,R}}]_{1}), ..., \boldsymbol{\alpha}([\boldsymbol{\phi}_{\text{M,R}}]_{L_{\text{M,R}}})\Big],
\end{equation}
\begin{equation}
    \bA(\boldsymbol{\phi}_{\text{M,R}}) = \Big[ \boldsymbol{\alpha}([\boldsymbol{\phi}_{\text{M,R}}]_{1}), ..., \boldsymbol{\alpha}([\boldsymbol{\phi}_{\text{M,R}}]_{L_{\text{M,R}}})\Big].
\end{equation}
Similarly, the RIS-BS channel $\bH_{\text{R,B}}$ is defined as
\begin{align}
    \label{eq:hbr}
    \bH_{\text{R,B}} &= \sum\limits_{l = 1}^{L_{\text{R,B}}} [\boldsymbol{\rho}_{\text{R,B}}]_l \boldsymbol{ \alpha}([\boldsymbol{\phi}_{\text{R,B}}]_l ) \boldsymbol{\alpha}^{\mathsf{H}}([\boldsymbol{\theta}_{\text{R,B}}]_l),\nonumber \\
    & = \bA(\boldsymbol{\phi}_{\text{R,B}})\mathrm{diag}(\boldsymbol{\rho}_{\text{R,B}})\bA^{\mathsf{H}}(\boldsymbol{\theta}_{\text{R,B}}),
\end{align}
where $[\boldsymbol{\rho}_{\text{R,B}}]_l$, $\boldsymbol {\alpha}([\boldsymbol{\phi}_{\text{R,B}}]_l )$, and $\boldsymbol{\alpha}([\boldsymbol{\theta}_{\text{R,B}}]_l)$ are the $l$th propagation path gain, and array response vectors as a function of $[\boldsymbol{\phi}_{\text{R,B}}]_l$ and $[\boldsymbol{\theta}_{\text{R,B}}]_l$, respectively. By applying~\eqref{eq:hrm} and~\eqref{eq:hbr}, and taking RIS into consideration, we can express the complete end-to-end MS-RIS-BS channel as 
\begin{equation}
\label{eq:hc}
      \bH= \bH_{\text{R,B}}\boldsymbol{\Omega}\bH_{\text{M,R}},
\end{equation}
where $\boldsymbol{\Omega}\in\mathbb{C}^{N_{\text{R}}\times N_{\text{R}}}$ is the phase control matrix at the RIS with constant-modules entries on the diagonal. By assuming that the RIS is composed by discrete phase shifters, the phase control matrix is defined as $\left[\mathbf{\Omega}\right]_{k,k}=\exp\left({j\omega_k}\right)$, where $\omega_k \in [0,2\pi)$. Moreover, we also define the effective channel $\bG$ as 
\begin{equation}
\label{eq:g}
    \bG  = \mathrm{diag} (\boldsymbol{\rho}_{\text{R,B}})\bA^ {\mathsf{H}}(\boldsymbol{\theta}_{\text{R,B}}){\boldsymbol{\Omega}}\bA(\boldsymbol{\phi}_{\text{M,R}})\mathrm{diag}(\boldsymbol{\rho}_{\text{M,R}}),     
\end{equation}
which depends on the phase control matrix and angular parameters associated with the RIS and propagation path gains. 

\section{Sounding Procedure for RIS architectures}
\label{sec:sound_procedure}
In this section, we detail the sounding procedure for the passive RIS architecture, used for comparison purposes, and the hybrid RIS, which is composed of both passive and active elements, as illustrated by Fig.~\ref{fig1:RIS}. 

\subsection{Passive RIS Architecture}
In the literature, the CE for the passive RIS architecture has been conducted in~\cite{he2020,ardah2020trice,CE_He_Zhen-Qing}. In this case, the CE of the individual channels or the cascaded one as in~\eqref{eq:hc} can be performed only at the BS or MS, since the RIS elements are passive without any baseband processing capability. In addition, the number of RIS elements is usually much larger than the number of antennas at BS and MS, which leads to very high-dimensional channel matrices/vectors. We aim at estimating the channel parameters in~\eqref{eq:hrm} and~\eqref{eq:hbr} other than the whole channel matrices by taking into consideration the inherent channel sparsity. 

We assume block fading channel, so that we can divide the coherence time into two sub-intervals, the first one for CE and the second one for data transmission (DT). We further divide the CE sub-interval into $K$ blocks. We assume the uplink pilot-based training procedure, where MS sends a series of training matrices $\bX_k$ during $k=1,\cdots,K$ to the BS. The signal is reflected at the RIS by $\mathbf{\Omega}_k$, combined\footnote{We consider analog combining matrix with random phases. The use of analog combining is a common assumption in mmWave MIMO systems.} at the BS by $\bW_k$, and received as ${\mathbf{Y}_{\mathrm{P}}}_{k}$. 
The received signal at the BS is
\begin{equation}
     \label{eq_y1}
    {\mathbf{Y}_{\mathrm{P}}}_{k}=\sqrt{P_{\text{T}}} \beta_{\text{2}}\mathbf{W}_{k}^{\mathsf{H}}\bH_{\text{R,B}}\boldsymbol{\Omega}_k\bH_{\text{M,R}}\bX_{k}+ \bW_{k}^{\mathsf{H}}\bZ_{k},          
    \mbox{for } k=1,\cdots,K,
\end{equation}
where $P_{\text{T}}$ is the transmit power, $\mathbf{Z}_k$ is the additive Gaussian noise with each entry distributed as $\mathcal{CN} (0,\sigma^2)$ and $\beta_{\text{2}} = \sqrt{\frac{1}{\beta(d_{\text{1}},d_{\text{2}})}}$, with $\beta(d_{\text{1}},d_{\text{2}})$ being the overall path loss in the MS-RIS-BS route and $d_{\text{1}}$, $d_{\text{2}}$ denoting the distance between MS and RIS and that between RIS and BS, respectively. 

\subsection{Hybrid RIS Architecture}
\begin{figure} [!t]
    \centering
    \includegraphics[width=0.7\columnwidth]{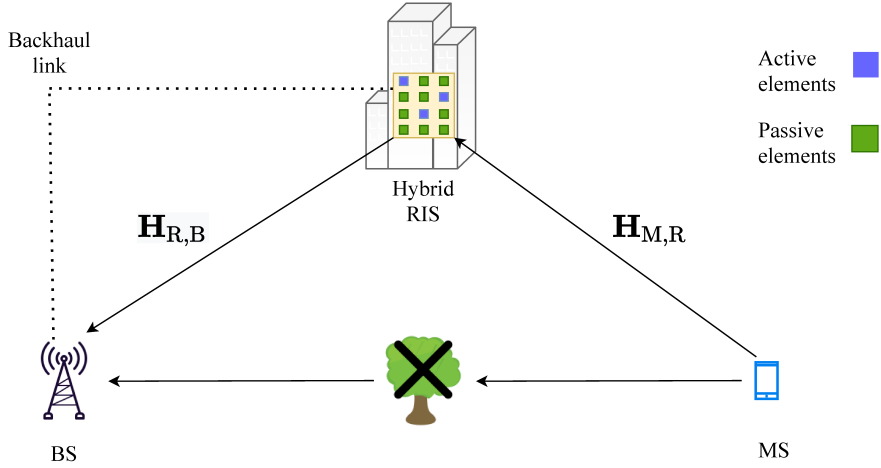}
    \caption{Hybrid RIS with both passive and active elements.}
    \label{fig1:RIS}
\end{figure}

Thanks to the availability of measurements/received signals at the active elements illustrated in the Fig.~\ref{fig1:RIS}, we can perform the CE at RIS in the hybrid architecture.
After the CE, the estimates of the channel parameters are transmitted to the BS by using the backhaul\footnote{It is possible to use high speed fronthaul link between RIS and BS in order to reduce the complexity of the feedback.}. Furthermore, by adopting the hybrid RIS architecture, we can decouple the CE problem  and further simplify the CSI acquisition~\cite{ActElements}. 
We assume $M$ out of $N_{\text{R}}$ RIS elements are active and $N_{\text{RF,R}} =  M$ RF chains at the RIS. We conduct uplink pilot training, where the MS sends the beam training matrix $\bX\in\mathbb{C}^{N_{\text{M}}\times T}$ to the BS and the RIS, where $T$ is the number of training sequences.  Similar to passive RIS sounding, we also divide the CE sub-interval into $K$ blocks, where each block has $T$ channel uses.

\begin{figure} [!t]
    \centering
    \includegraphics[width=0.6\columnwidth]{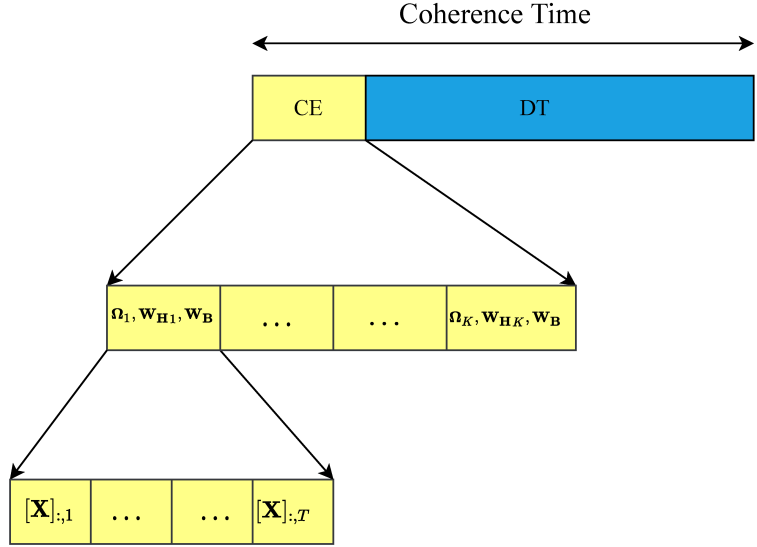}
    \caption{Uplink training procedure for the hybrid RIS. We divide the coherence time in two sub-intervals, one for CE and the other for DT. We further divide the CE sub-interval into $K$ blocks with each block containing $T$ channel uses.}
    \label{fig:training}
\end{figure}

The pilot signals are received at RIS by the $M$ active elements, indexed by the set $\mathbb{M}$, i.e., $|\mathbb{M}| = M$, while the remaining passive elements only reflect the incident signal with the phase control matrix $\boldsymbol{\Omega}_k$. It is worth mentioning that $\boldsymbol{\Omega}_{k}$ differs from the phase control matrix in the passive RIS since $[\boldsymbol{\Omega}_{k}]_{i,i} = 0$ for $i \in \mathbb{M}$. After the reflection at the RIS, the signal is further combined at the BS by $\mathbf{W}_{\text{B}} \in\mathbb{C}^{N_{\text{B}} \times N_{\text{C,B}}}$, and received as $\mathbf{Y}_{k}$, where $N_{\text{C,B}}$ is the number of columns of the combining matrix. 
In the training procedure, we keep the combining matrix $\mathbf{W}_{\text{B}}$ and the training matrix $\bX$ constant over all the blocks, while the phase control matrix $\boldsymbol{\Omega}_{k}$ at the RIS varies from block to block. Fig.~\ref{fig:training} summarizes the training procedure for the hybrid RIS, occurring prior to DT. The received signal at the RIS ${\mathbf{Y}_{\text{H}}}_{k}$ is expressed as
\begin{equation}
\label{eq:r_s_ris_block}
      {\mathbf{Y}_{\text{H}}}_{k} =\sqrt{P_{\text{T}}} \beta_{\text{1}}{\mathbf{W}_{\text{H}}}_k\mathbf{H}_{\text{M,R}}\mathbf{X}+{\mathbf{W}_{\text{H}}}_k{\mathbf{Z}_{1}}_k, \mbox{for } k=1,\cdots,K,
\end{equation}
where ${\mathbf{W}_{\text{H}}}_{k}$ is a row-selection matrix containing $M$ rows of a $N_{\text{R}} \times N_{\text{R}}$ identity matrix, ${\mathbf{Z}_{1}}_k\in \mathbb{C}^{M\times T}$ is the Gaussian noise with each entry distributed as $\mathcal{CN} (0,\sigma^2)$. The term $\beta_{\text{1}}$ is expressed as $\beta_{\text{1}} =\sqrt{\frac{1}{\beta(d_{\text{1}})}}$, where $\beta(d_{\text{1}})$ is the path loss associated with the MS-RIS channel. Moreover, the received signal at BS is expressed as
\begin{equation}
    \label{eq:r_s_bs_block}
    \mathbf{Y}_k = \sqrt{P_{\text{T}}} \beta_{\text{2}}\mathbf{W}_{\text{B}}^{\mathsf{H}}\mathbf{H}_{\text{R,B}}\boldsymbol{\Omega}_k\mathbf{H}_{\text{M,R}}\mathbf{X}+ \mathbf{W}_{\text{B}}^{\mathsf{H}}{\mathbf{Z}_{2}}_k,
\end{equation}
where ${\mathbf{Z}_{2}}_k\in \mathbb{C}^{N_{\text{B}}\times T}$ is the Gaussian noise also with each entry distributed as $\mathcal{CN} (0,\sigma^2)$. After $K$ blocks, the complete received signal at the RIS is $\mathbf{Y}_{\text{H}}  = [{\bY_{\text{H}}}_1^{\mathsf{T}},\cdots,{\bY_{\text{H}}}_K^{\mathsf{T}}]^{\mathsf{T}}
\in \mathbb{C}^{M K\times T}$, which can be summarized as 
\begin{equation}
        \label{eq:rris}
        \mathbf{Y}_{\text{H}} = \sqrt{P_{\text{T}}} \beta_{\text{1}}\mathbf{W}_{\text{H}}\mathbf{H}_{\text{M,R}}\mathbf{X}+ \bar{\mathbf{Z}}_{1}, 
\end{equation}
where $\bW_{\text{H}} = [{\bW_{\text{H}}}_1^{\mathsf{T}},\cdots,{\bW_{\text{H}}}_K^{\mathsf{T}}]^{\mathsf{T}}\in \mathbb{C}^{M K\times N_{\text{R}}}$, $\bar{\bZ}_{1} = [({\bW_{\text{H}}}_1{\bZ_{1}}_1)^{\mathsf{T}},\cdots,({\bW_{\text{H}}}_K{\bZ_{1}}_K)^{\mathsf{T}}]^{\mathsf{T}}\in \mathbb{C}^{M K \times T}$. The complete received signal at the BS $\mathbf{Y} = [\mathbf{Y}_1,\cdots,\mathbf{Y}_K]\in \mathbb{C}^{N_{\text{C,B}}\times T K}$ is  defined as
\begin{equation}
 \label{eq:bs}
        \mathbf{Y} = \sqrt{P_{\text{T}}} \beta_{\text{2}}\mathbf{W}_{\text{B}}^{\mathsf{H}}\mathbf{H}_{\text{R,B}}\mathbf{U}+ \bar{\mathbf{Z}}_{2}, 
\end{equation}
where $\bU = [\boldsymbol{\Omega}_{1}\bH_{\text{M,R}}\bX,\cdots,\boldsymbol{\Omega}_{K}\bH_{\text{M,R}}\bX] \in \mathbb{C}^{N_{\text{R}}\times T K}$ and $\bar{\bZ}_{2}= [\bW_{\text{B}}^{\mathsf{H}}{\bZ_{2}}_1,\cdots,\bW_{\text{B}}^{\mathsf{H}}{\bZ_{2}}_K]\in \mathbb{C}^{N_{\text{C,B}}\times T K }$.

We propose a two-stage CE to recover the channel parameters from the received signals in~\eqref{eq:rris} and~\eqref{eq:bs}. At the first stage, we recover the channel parameters in  $\mathbf{H}_{\text{M,R}}$ based on the received signal at the RIS in~\eqref{eq:rris}. After the estimation of $\mathbf{H}_{\text{M,R}}$, the RIS sends the estimates of the channel parameters to the BS via the backhaul link. The BS reconstructs the channel $\mathbf{H}_{\text{M,R}}$ based on the estimates of the channel parameters, denoted as $\hat{\mathbf{H}}_{\text{M,R}}$. At the second stage, we assume $\hat{\mathbf{H}}_{\text{M,R}}\approx \mathbf{H}_{\text{M,R}}$ and target at the recovery of the channel parameters in $\mathbf{H}_{\text{R,B}}$ from the received signal at the BS in~\eqref{eq:bs}. 

\section{RIS Channel Estimation}
\label{section: CE}
The channel estimation for mmWave MIMO systems have been investigated by applying on-the-grid compressive sensing (CS) methods~\cite{lee2016channel} as well as more advanced off-the-grid CS methods, such as ANM~\cite{harnessing,yang2015gridless}. 

We resort to the ANM for the recovery of the angular parameters or equivalent spatial frequencies. We substitute the angular parameters with spatial frequencies in the following. Thus, we can express~\eqref{eq:hrm} as 
\begin{equation}
\bH_{\text{M,R}}= \sum\limits_{l = 1}^{L_{\text{M,R}}} [\boldsymbol{\rho}_{\text{M,R}}]_l \boldsymbol{\alpha}([\mathbf{g}_{1}]_l ) \boldsymbol{\alpha}^{\mathsf{H}}([\mathbf{f}_{1}]_l), 
\end{equation}
where the spatial frequencies $\bf_{1} = \sin (\boldsymbol{\theta}_{\text{M,R}})$, $\bg_{1} = \sin (\boldsymbol{\phi}_{\text{M,R}})$. With the abuse of $\boldsymbol{\alpha}(\cdot)$, we redefine $[\boldsymbol{ \alpha}([\mathbf{g}_{1}]_l )]_{n} =  \exp\{
j\pi(n-1)[\mathbf{g}_{1}]_l\}$.  Similarly, we express~\eqref{eq:hbr} as
\begin{equation}
\bH_{\text{R,B}}= \sum\limits_{l = 1}^{L_{\text{R,B}}} [\boldsymbol{\rho}_{\text{R,B}}]_l \boldsymbol{\alpha}([\mathbf{g}_{2}]_l ) \boldsymbol{\alpha}^{\mathsf{H}}([\mathbf{f}_{2}]_l), 
\end{equation}
where the spatial frequencies $\bf_{2} = \sin (\boldsymbol{\theta}_{\text{R,B}})$, $\bg_{2} = \sin (\boldsymbol{\phi}_{\text{R,B}})$. These spatial frequencies are within $[0,1)$ by assuming the angles are within $[0,\pi)$. The atomic set of $\bH_{\text{M,R}}$, denoted by $\mathcal{A}_{\text{M}}$, is expressed as 
\begin{equation}
    \mathcal{A}_{\text{M}} = \left \{ \bQ_{1}(f_{1},g_{1}): f_{1}\ \in \Big[ 0,1 \Big),  g_{1}\in \Big[ 0,1 \Big) \right \},
\end{equation}
where $\bQ_{1}(f_{1},g_{1}) = \boldsymbol{\alpha}(f_{1}) \boldsymbol{\alpha}^{\mathsf{H}}(g_{1})$ is the matrix atom. Similarly, the atomic set of $\bH_{\text{R,B}}$ is given by 
\begin{equation}
    \mathcal{A}_{\text{N}} = \left \{ \bQ_{2}(f_{2}, g_{2}): f_{2}\ \in \Big[ 0,1 \Big),  g_{2}\in \Big[ 0,1 \Big) \right \},
\end{equation}
where $\bQ_{2}(f_{2}, g_{2}) = \boldsymbol{\alpha}(f_{2}) \boldsymbol{\alpha}^{\mathsf{H}}(g_{2})$ denotes the matrix atom.

\subsection{First Stage of CE for Hybrid RIS}
At the first stage, we aim at the recovery of the channel parameters in $\bH_{\text{M,R}}$. For this reason, we formulate the atomic norm with respect to the atomic set $\mathcal{A}_{\text{M}}$ as
\begin{equation}
    \|\bH_{\text{M,R}}\|_{\mathcal{A}_{\text{M}}} = \inf \{q: \bH_{\text{M,R}} \in \mathrm{conv}({\mathcal{A}_{\text{M}}}) \} .
\end{equation}
The equivalent form as a semidefinite programming (SDP) problem is~\cite{tsai2018millimeter}
\begin{equation}
\begin{split}
    \|\mathbf{H}_{\text{M,R}}\|_{\mathcal{A}_{\text{M}}} &= \mathrm{inf}_{\{\mathbf{C},\mathbf{V}\}} \Big\{\frac{1}{2 N_\text{M}} \mathrm{Tr}(\mathbb{T}(\mathbf{C})) + \frac{1}{2 N_\text{R}} \mathrm{Tr}(\mathbb{T}(\mathbf{V})) \Big\} \\
    &\textrm{s.t} \quad \begin{bmatrix}
        \mathbb{T}(\mathbf{C}) & \mathbf{H}_{\text{M,R}}\\
        \mathbf{H}_{\text{M,R}}^{\mathsf{H}} & \mathbb{T}(\mathbf{V})
    \end{bmatrix}
    \succeq 0,
\end{split}
\end{equation}
where $\mathbb{T}(\mathbf{C})$ and $\mathbb{T}(\mathbf{v})$ are 2-level Toeplitz matrices. The recovery of the angles $\boldsymbol{\theta}_{\text{M,R}}$ and $\boldsymbol{\phi}_{\text{M,R}}$ (or equivalently $\bf_1$ and $\bg_1$) can be done by addressing the following convex problem
\begin{equation}
    \label{eq:anm_h1}
    \hat{\bH}_{\text{M,R}}= \arg \min_{\bH_{\text{M,R}}} \tau \|\bH_{\text{M,R}}\|_{\mathcal{A}_{\text{M}}} + \frac{1}{2} \|\sqrt{P_{\text{T}}}  \beta_{\text{1}}\mathbf{W}_{\text{H}}\bH_{\text{M,R}}\bX -  \bY_{\text{H}}\|^{2}_{\mathrm{F}},
\end{equation}
where $\tau$ is the regularization parameter set as $\tau \varpropto \sigma\sqrt{N_\text{R} N_\text{M}\log (N_\text{R} N_\text{M}})$. Using the SDP formulation, we further formulate the problem as  
\begin{equation}
\begin{split}
\label{eq_anm_stage1}
    \hat{\mathbf{H}}_{\text{M,R}} = \arg \min_{\mathbf{H}_{\text{M,R}},\mathbf{{C}},\mathbf{{V}}} \quad &\frac{\tau}{2 N_\text{M}} \mathrm{Tr}(\mathbb{T}(\mathbf{C})) +  \frac{\tau}{2 N_\text{R}} \mathrm{Tr}(\mathbb{T}(\mathbf{V})) \\
    &+ \frac{1}{2} \|\sqrt{P_{\text{T}}} \beta_{\text{1}} \mathbf{W}_{\text{H}}\mathbf{H}_{\text{M,R}}\mathbf{X} - \bY_{\text{H}}\|^{2}_{\mathrm{F}} \\
    \textrm{s.t} \qquad &\begin{bmatrix}
    \mathbb{T}(\mathbf{C}) & \mathbf{H}_{\text{M,R}}\\
    \mathbf{H}_{\text{M,R}}^{\mathsf{H}} &  \mathbb{T}(\mathbf{V})\\
    \end{bmatrix}
    \succeq 0.
\end{split}
\end{equation}
The solutions of the Toeplitz matrices $\mathbb{T}(\mathbf{C})$ and $\mathbb{T}(\mathbf{V})$ lead us to the recovery of the angles $\boldsymbol{\theta}_{\text{M,R}}$ and $\boldsymbol{\phi}_{\text{M,R}}$, respectively, by applying the ROOTMUSIC algorithm~\cite{MUSIC}.\footnote{The ROOTMUSIC algorithm can be easily implemented by using the function \textit{rootmusic} in MATLAB.} We assume the order information of the angles and the number of paths as known prior information. We estimate the path gain vector $\boldsymbol{\rho}_{\text{M,R}}$ by applying the least squares (LS), which results in
\begin{equation}
    \label{eq_path_gain}
    \hat{\boldsymbol{\rho}}_{\text{M,R}} = 
   \Big[\sqrt{P_{\text{T}}}\beta_{\text{1}}(\mathbf{X}^{\mathsf{T}}\otimes \mathbf{W}_{\text{H}})\big((\bA^*(\hat{\boldsymbol{\theta}}_{\text{M,R}}) \odot \bA(\hat{\boldsymbol{\phi}}_{\text{M,R}})\big)\Big]^{\dagger}\mathbf{y}_{\text{H}},
\end{equation}
where $\mathbf{y}_{\text{H}}$ is defined as $\mathbf{y}_{\text{H}} = \mathrm{vec}({\mathbf{Y}_{\text{H}}})$.

\subsection{Second Stage of CE for Hybrid RIS}
At the second stage, we aim at extracting the channel parameters from the received signals at BS. In order to simplify the CE, we use the estimate of  $\hat{\mathbf{H}}_{\text{M,R}}$ from the first stage, which results in 
\begin{equation}
    \label{eq:hatU}
    \hat{\bU}=[\boldsymbol{\Omega}_{1}\hat{\mathbf{H}}_{\text{M,R}}\mathbf{X},\cdots,\boldsymbol{\Omega}_{K}\hat{\mathbf{H}}_{\text{M,R}}\mathbf{X}] \in \mathbb{C}^{N_{\text{R}}\times TK}.
\end{equation}
We formulate the atomic norm of $\bH_{\text{R,B}}$ as 
\begin{equation}
    \|\bH_{\text{R,B}}\|_{\mathcal{A}_{\text{N}}} = \inf \{q: \bH_{\text{R,B}} \in \mathrm{conv}({\mathcal{A}_{\text{N}}}) \} .
\end{equation}
By following the SDP formulation, we can write 
\begin{equation}
\begin{split}
    \|\mathbf{H}_{\text{R,B}}\|_{\mathcal{A}_{\text{N}}} &= \mathrm{inf}_{\{\mathbf{S},\mathbf{O}\}} \Big\{\frac{1}{2 N_\text{R}} \mathrm{Tr}(\mathbb{T}(\mathbf{S})) + \frac{1}{2 N_\text{B}} \mathrm{Tr}(\mathbb{T}(\mathbf{O})) \Big\} \\
    &\textrm{s.t} \qquad \begin{bmatrix}
        \mathbb{T}(\mathbf{S}) & \mathbf{H}_{\text{R,B}}\\
        \mathbf{H}_{\text{R,B}}^{\mathsf{H}} & \mathbb{T}(\mathbf{O})
    \end{bmatrix}
    \succeq 0.
\end{split}
\end{equation}
where $\mathbb{T}(\mathbf{S})$ and $\mathbb{T}(\mathbf{O})$ are 2-level Toeplitz matrices. In order to recover the angles $\boldsymbol{\theta}_{\text{R,B}}$ and $\boldsymbol{\phi}_{\text{R,B}}$ in $\bH_{\text{R,B}}$, we formulate the problem as 
\begin{equation}
    \hat{\bH}_{\text{R,B}}= \arg \min_{\bH_{\text{R,B}}} \nu \|\bH_{\text{R,B}}\|_{\mathcal{A}_{\text{N}}} + \frac{1}{2} \|\sqrt{P_{\text{T}}}\beta_{\text{2}} \mathbf{W}_{\text{B}}^{\mathsf{H}}\mathbf{H}_{\text{R,B}}\hat{\bU} - \bY\|^{2}_{\mathrm{F}},
\end{equation}
where the regularization parameter $\nu$ is set as $\nu\varpropto \sigma\sqrt{N_\text{R} N_\text{B}\log (N_\text{R} N_\text{B}})$. By following the SDP formulation, we define the problem as 
\begin{equation}
\begin{split}
\label{eq_anm_stage2}
    \hat{\mathbf{H}}_{\text{R,B}} = \arg \min_{\bH_{\text{R,B}},\mathbf{{S}},\mathbf{{O}}} \quad &\frac{\nu}{2 N_\text{R}} \mathrm{Tr}(\mathbb{T}(\mathbf{S})) +  \frac{	\nu}{2 N_\text{B}} \mathrm{Tr}(\mathbb{T}(\mathbf{O})) \\
    &+ \frac{1}{2} \|\sqrt{P_{\text{T}}} \beta_{\text{2}} \mathbf{W}_{\text{B}}^{\mathsf{H}}\mathbf{H}_{\text{R,B}}\hat{\bU} - \bY\|^{2}_{\mathrm{F}} \\
    \textrm{s.t} \qquad & \begin{bmatrix}
    \mathbb{T}(\mathbf{S}) & \mathbf{H}_{\text{R,B}}\\
    \mathbf{H}_{\text{R,B}}^{\mathsf{H}} &  \mathbb{T}(\mathbf{O})\\
    \end{bmatrix}
    \succeq 0.    
\end{split}
\end{equation}
Similarly, the angles $\boldsymbol{\theta}_{\text{R,B}}$ and $\boldsymbol{\phi}_{\text{R,B}}$ can be recovered based on the solutions of $\mathbb{T}(\mathbf{S})$ and $\mathbb{T}(\mathbf{O})$, respectively, which can be solved by ROOTMUSIC algorithm~\cite{MUSIC}. The estimate of the path gain vector is addressed by LS as
\begin{equation}
    \label{eq_path_gain2}
    \hat{\boldsymbol{\rho}}_{\text{R,B}} = 
   \Big[\sqrt{P_{\text{T}}}\beta_{\text{2}}(\hat{\mathbf{U}}^{\mathsf{T}}\otimes \mathbf{W}_{\text{B}}^{\mathsf{H}})\big((\bA^*(\hat{\boldsymbol{\theta}}_{\text{R,B}}) \odot \bA(\hat{\boldsymbol{\phi}}_{\text{R,B}})\big)\Big]^{\dagger}\mathbf{y},
\end{equation}
where $\mathbf{y} = \mathrm{vec}({\mathbf{Y}})$. The proposed two-stage CE method is summarized in the \textbf{Algorithm 1}.

With the estimation of all the channel parameters, we can calculate the angle differences associated with the RIS (written as a matrix) and the products of path gains (written as a vector). The angle differences are functions of the estimates of $\boldsymbol{\phi}_{\text{R,B}}$ and $\boldsymbol{\theta}_{\text{M,R}}$, expressed as
\begin{align}
\label{eq_ang_diff}
    [\hat{\boldsymbol{\Delta}}]_{lp} &=  \mathrm{asin} \big[ \sin{([\hat{\boldsymbol{\phi}}_{\text{M,R}}]_{l})} - \sin{([\hat{\boldsymbol{\theta}}_{\text{R,B}}]_{p})} \big],\nonumber\\
    &\text{for}\; l = 1,\cdots, L_{\text{M,R}}, p = 1,\cdots, L_{\text{R,B}}. 
\end{align}
Moreover, the estimated products of path gains $\hat{\boldsymbol{\rho}} \in \C^{L_{\text{R,B}}L_{\text{M,R}} \times 1}$ are defined as follows 
\begin{equation}
\label{eq_prod_path}
    \hat{\boldsymbol{\rho}} = \hat{\boldsymbol{\rho}}_\text{R,B} \otimes \hat{\boldsymbol{\rho}}_\text{M,R}.
\end{equation}

The training overhead for the proposed CE for hybrid RIS-aided mmWave MIMO systems is given as
\begin{equation}
    T_\text{H} =  K T \Big\lceil\frac{N_{\text{C,B}}}{N_{\text{RF,B}}}\Big\rceil
    \Big\lceil\frac{M}{N_{\text{RF,R}}}\Big\rceil,
\end{equation}
where $N_{\text{RF,B}}$ is the number of RF chains at BS.

\begin{algorithm}
    \SetAlgoLined
    \SetKwInOut{Input}{Input}
    \SetKwInOut{Output}{Output}
    \Input{$\bf_{1}$, $\bg_{1}$, $\bf_{2}$, $\bg_{2}$, $\boldsymbol{\rho}_\text{R,B}$, $\boldsymbol{\rho}_\text{M,R}$, $M$, $K$}
    Define $\bH_{\text{M,R}}$, $\bH_{\text{R,B}}$, $\mathbf{W}_{\text{B}}$, and $\bX$; \\   
    \Comment*[h]{Uplink training} \\
    \For{$k\gets1$ \KwTo $K$}{
    Define $|\mathbb{M}| = M$; \\
    Compute ${\mathbf{Y}_{\text{H}}}_{k}$ accordingly to~\eqref{eq:r_s_ris_block}; \\
    Set $[\boldsymbol{\Omega}_{k}]_{i,i} = 0$, for $i \in \mathbb{M}$;  \\
    Compute $\mathbf{Y}_{k}$ by following~\eqref{eq:r_s_bs_block};
    }
    Collect the received signals  $\bY_{\text{H}}=[{\bY_{\text{H}}}_1^{\mathsf{T}},\cdots,{\bY_{\text{H}}}_K^{\mathsf{T}}]^{\mathsf{T}}\in \mathbb{C}^{M K\times T}$ and $\bY = [\bY_1,\cdots,\bY_K]\in \mathbb{C}^{N_{\text{C,B}}\times T K}$; \\
    \Comment*[h]{First stage of CE} \\
    Estimate $\boldsymbol{\theta}_{\text{M,R}}$ and $\boldsymbol{\phi}_{\text{M,R}}$ by following~\eqref{eq_anm_stage1};\\ 
    Estimate $\boldsymbol{\rho}_{\text{M,R}}$ by using LS~\eqref{eq_path_gain};\\
    \Comment*[h]{Second stage of CE} \\
    Reconstruct $\hat{\bH}_{\text{M,R}}$ at the BS;  \\
    Compute $\hat{\bU}$ by following~\eqref{eq:hatU}; \\
    Estimate $\boldsymbol{\theta}_{\text{R,B}}$ and $\boldsymbol{\phi}_{\text{R,B}}$ by following~\eqref{eq_anm_stage2};\\
    Estimate $\boldsymbol{\rho}_{\text{R,B}}$ by applying~\eqref{eq_path_gain2};\\
    Reconstruct $\hat{\bH}_{\text{R,B}}$; \\
    \Output{$\hat{\bH}_{\text{M,R}}$, $\hat{\bH}_{\text{R,B}}$, $\hat{\boldsymbol{\rho}}_\text{M,R}$, $\hat{\boldsymbol{\theta}}_\text{M,R}$, $\hat{\boldsymbol{\phi}}_\text{M,R}$, $\hat{\boldsymbol{\rho}}_\text{R,B}$, $\hat{\boldsymbol{\theta}}_\text{R,B}$, and $\hat{\boldsymbol{\phi}}_\text{R,B}$}
    \caption{The Proposed Two-Stage CE for Hybrid RIS}
\end{algorithm}

\subsection{Passive RIS CE}
The CE for the passive RIS-aided mmWave MIMO systems is addressed using the two-stage procedure via ANM, detailed in~\cite{he2020anm}. We summarize the CE procedures for this architecture here as well for the sake of completeness.  At the first stage, we estimate AoAs at the BS ($\boldsymbol{\phi}_{\text{R,B}}$) and AoDs at the MS ($\boldsymbol{\theta}_{\text{M,R}}$) using the multiple measurement vectors (MMV) model. We design the beam training matrix at MS and the combining matrix at the BS based on the estimates from the first stage. At the second stage, we recover the angle differences and products of path gains, denoted by $\boldsymbol{\Delta}$ and $\boldsymbol{\rho}$, respectively. 

The training overhead for the passive RIS CE via ANM is 
\begin{equation}
        T_{\text{P}} = N_{0}
        \ceil[\Big]{
        \frac{M_0}{N_{\text{RF},\text{B}}}} + T L_{\text{M,R}}\ceil[\Big]{
        \frac{L_{\text{R,B}}}{N_{\text{RF},\text{B}}}},
\end{equation}
where $N_{0}$ is the number of training beams at the first stage of CE and $M_0$ is the number of columns of the combining matrix at the BS~\cite{he2020anm}. 

\section{Design of RIS Phase Control Matrix and Beamforming Vectors}
\label{sec:beamforming}
In this section, we present the design of the phase control matrix at the RIS and the beamforming vectors at BS and MS. We clarify that we follow the same procedure for both RIS architectures, i.e., the hybrid and passive RISs. 

\subsection{RIS Phase Control Matrix}
We design the RIS phase control matrix at the RIS based on the maximization of the power of the effective channel, as defined in~\eqref{eq:g}. Based on this criterion, the optimal phase control matrix is
\begin{equation}
    \label{eq:G1}
    \boldsymbol{\Omega}^{*}  = \arg \max_{\boldsymbol{\Omega}} \|\bG\|^{2}_{\mathrm{F}},
\end{equation}
where $\|\bG\|^{2}_{\mathrm{F}}  =  \|\mathrm{diag}\small(\boldsymbol{\hat{\rho}}_{\text{R,B}}\small)\bA^ {\mathsf{H}}\small(\boldsymbol{\hat{\theta}}_{\text{R,B}}\small){\boldsymbol{\Omega}}\bA\small(\boldsymbol{\hat{\phi}}_{\text{M,R}}\small)\mathrm{diag}\small(\boldsymbol{\hat{\rho}}_{\text{M,R}}\small)\|^{2}_{\mathrm{F}}$, and its $(l,p)$th entry can be written as
\begin{align}
    \label{eq:def_g_f}
        [\bG]_{l,p}  &= [\boldsymbol{\hat{\rho}}_{\text{R,B}}]_{p}\boldsymbol{\omega}^{\mathsf{T}}\boldsymbol{\alpha}([\hat{\boldsymbol{\Delta}}]_{l,p})[\boldsymbol{\hat{\rho}}_{\text{R,M}}]_{l},\nonumber\\
    &\text{for}\; l = 1,\cdots, L_{\text{R,B}}, p = 1,\cdots, L_{\text{R,M}},
\end{align} 
where $\boldsymbol{\Omega} = \mathrm{diag}(\boldsymbol{\omega})$. We further define the vectorization of $\bG$ as $\mathbf{g}=\mathrm{vec}(\bG)$, and the $i$th element of $\mathbf{g}$ is expressed as 
\begin{equation}
  \label{eq:vec_G}
    [\mathbf{g}]_{i} = [\hat{\boldsymbol{\rho}}]_{i}\boldsymbol{\omega} ^{\mathsf{T}}\boldsymbol{\alpha}([\hat{\boldsymbol{\delta}}]_{i}),\;
    \mbox{for }  i=1,\cdots,L_{\text{R,B}}L_{\text{M,R}},
\end{equation}
where $\hat{\boldsymbol{\delta}}=\mathrm{vec}(\hat{\boldsymbol{\Delta}})$.

By following~\eqref{eq:def_g_f} and~\eqref{eq:vec_G}, we can further express $\|\bG\|^{2}_{\mathrm{F}}$  as 
\begin{equation}
          \label{eq:g3}
          \|\bG\|^{2}_{\mathrm{F}} = \sum^{L_{\text{R,B}}L_{\text{M,R}}}_{i=1} |[\hat{\boldsymbol{\rho}}]_{i}\boldsymbol{\omega}^{\mathsf{T}}\boldsymbol{\alpha}([\hat{\boldsymbol{\delta}}]_{i})|^2.
\end{equation}
Substituting~\eqref{eq:g3} in~\eqref{eq:G1}, we define the optimal $\boldsymbol{\omega}^{*}$ as
\begin{equation}
    \begin{split}
    \label{eq: omega}
    \boldsymbol{\omega}^{*} & = \arg \max_{\boldsymbol{\omega}} \quad \displaystyle \sum^{L_{\text{R,B}}L_{\text{M,R}}}_{i=1} |[\hat{\boldsymbol{\rho}}]_{i}\boldsymbol{\omega} ^{\mathsf{T}}\boldsymbol{\alpha}([\hat{\boldsymbol{\delta}}]_{i})|^2 \\
    &= \arg \max_{\boldsymbol{\omega}} \quad \boldsymbol{\omega}^{\mathsf{T}}\bC \bC^{\mathsf{H}} \boldsymbol{\omega}^{*}, \\
    \end{split}    
\end{equation}
where
\begin{equation}
\mathbf{C}=\big[\boldsymbol{\alpha}([\boldsymbol{\hat{\delta}}]_{1}),\dots,\boldsymbol{\alpha}([\boldsymbol{\hat{\delta}}]_{L_{\text{R,B}}L_{\text{M,R}}})\big]\mathrm{diag}(\big[[\boldsymbol{\hat{\rho}}]_{1},\dots,[\hat{\boldsymbol{\rho}}]_{L_{\text{R,B}}L_{\text{M,R}}}\big]).
\end{equation}
We apply singular value decomposition (SVD) on $\bC\bC^{\mathsf{H}}$, resulting in $\bC\bC^{\mathsf{H}} = \bE\bD\bE^{\mathsf{H}}$. Then, in order to obtain the optimal $\boldsymbol{\omega}^*$, we select the first column of $\bE$ and further project it to the unit-modulus vector space, resulting in $\boldsymbol{\omega}^{*} = \exp (-j \mathrm{phase}([\bE]_{:,1}))$, where $\mathrm{phase}()$ means the element-wise operation of extracting the phase of the argument. 

\subsection{Beamforming Vectors}

The design of the beamforming vectors is based on $\hat{\bH} = \hat{\bH}_{\text{R,B}}\mathrm{diag}(\boldsymbol{\omega}^*)\hat{\bH}_{\text{M,R}}$. We conduct SVD on $\hat{\mathbf{H}}$, which results in $\hat{\bH} = \bU\bS\bV$, so that we select the first  left-singular vector and right-singular vector as  beamforming vectors, i.e., $\mathbf{f}=[\bV]_{:,1}$ and $\mathbf{w} = [\bU]_{:,1}$.

\section{CRLB Analyses}
\label{section: crlb}
The CRLB is the lower bound of any unbiased estimator~\cite{kay1993fundamentals}, so that it serves as a performance indicator of our proposed CE method. Thus, in this section, we develop the CRLB for the channel parameter estimation based on~\cite{kay1993fundamentals,pillutla2017bayesian}. Through our analyses, the observation noise in~\eqref{eq:rris} and~\eqref{eq:bs} are assumed to be Gaussian. In our proposed two-stage CE, we first estimate the channel $\bH_{\text{M,R}}$ based on the received signals at RIS, and then estimate the channel $\bH_{\text{R,B}}$ based on the received signals at BS. In order to align with our proposed CE procedure, we also divide the calculations of the CRLB into two steps.


Then, at the first stage of the proposed CE, the parameters $\boldsymbol{\zeta} = \{\boldsymbol{\theta}_{\text{M,R}},\boldsymbol{\phi}_{\text{M,R}},\boldsymbol{\rho}_{\text{M,R}} \}$ are estimated, which are related to the MS-RIS channel. The MSEs of the channel parameters in $\boldsymbol{\zeta}$ are lower bounded by $\text{CRLB}(\boldsymbol{\zeta}) = \bJ^{-1}(\boldsymbol{\zeta})$,
where $\bJ(\boldsymbol{\zeta})$ is the fisher information matrix (FIM) of $\boldsymbol{\zeta}$. The details of the derivations of the FIM are provided in Appendix~\ref{appendix:CRLB1}.

Next, at the second stage of the proposed CE procedure, the parameters related to the RIS-BS channel $\boldsymbol{\eta} = \{ \boldsymbol{\theta}_{\text{R,B}},\boldsymbol{\phi}_{\text{R,B}}, \boldsymbol{\rho}_{\text{R,B}}\}$ are estimated. 
For simplicity, we calculate the CRLB for $\boldsymbol{\eta}$ by considering the following approximation,
\begin{equation}
    \label{eq_HRM}
    \hat{\bH}_{\text{M,R}} \approx \bH_{\text{M,R}}, 
\end{equation}
which leads to
\begin{equation}
    \tilde{\bU} = [\boldsymbol{\Omega}_{1}\mathbf{H}_{\text{M,R}}\mathbf{X},\cdots,\boldsymbol{\Omega}_{K}\mathbf{H}_{\text{M,R}}\mathbf{X}].
\end{equation}
Finally, the details of the derivations of $\text{CRLB}(\boldsymbol{\eta}) = \bJ^{-1}(\boldsymbol{\eta})$ are included in Appendix~\ref{appendix:CRLB2}.

\begin{figure} 
    \centering
    \includegraphics[width=0.6\columnwidth]{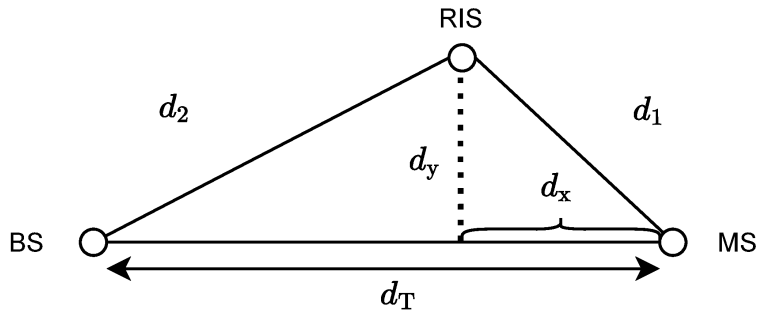}
    \caption{Network topology.}
    \label{fig:path_loss}
\end{figure}

\section{Performance Evaluation}
\label{section: perfomance}
In this section, we describe the parameter setup and the performance metrics to evaluate the proposed two-stage CE approach. The propagation path gains are distributed as $\mathcal{CN} (0,1)$. We assume $N_{\text{B}} = 16$, $N_{\text{R}} \in \{ 32, 64 \}$,  $N_{\text{M}}= 16$, $L_{\text{R,B}} = L_{\text{M,R}} = 2$. We perform 1000 trials to average out the results. As shown in Fig.~\ref{fig:path_loss}, we consider the BS at a fixed location~(0,0), the RIS is at the location~($x$,$y$) and the MS at~($x_{T}$,0), where $x = d_{\text{T}}-d_{\text{x}}$, $y= d_\text{y}$ and $x_{T}=d_{\text{T}}$. As a consequence, the distance between MS and RIS is $d_{\text{1}} = {\sqrt{d_{\text{x}}^2 + d_{\text{y}}^2}}$, while the distance between BS and RIS is $d_{\text{2}} = {\sqrt{(d_{\text{T}} - d_{\text{x}})^2 + d_{\text{y}}^2}}$. The path loss as a function of distance is given by~\cite{zhang2020capacity}, so that $\boldsymbol{\beta}(d_{\text{1}}) =\beta_\text{0}\Big(\frac{d_{0}}{d_{\text{1}}} \Big)^{\gamma}$ and $\beta(d_{\text{2}}) = \beta_\text{0}\Big(\frac{d_{0}}{d_{\text{1}}d_{\text{2}}} \Big)^{\gamma}$,
where $\beta_\text{0} = (\frac{\lambda}{4 \pi d_{0}})^{2}$ is the path loss in the reference distance $d_{0}$, $\lambda$ is denotes the wavelength, defined as the ratio between the $c$ speed of the light in meters per second ($c=3\times10^{8}$) and $f_{c}$ the carrier frequency in Hz. We set  $d_{0}$ as $1$ [m], the path loss exponent as $\gamma=3$, $f_{c}=28$ [MHz], the noise power density as $N_0 = -173$ [dBm/Hz], the bandwidth as $B = 100$ [MHz], and the transmit power as $P_{\text{T}} = 0:5:20$ [dBm]. Table~\ref{tab1:Setups2} summarizes the parameters of the system model. 
\begin{table}
    \centering
   \caption{Parameter Setup.}
    \label{tab1:Setups2}
    \begin{tabular}{clllllllc}
        \toprule
       Symbol & Parameter  & Value \\
       \midrule
        $\gamma$ & Path loss exponent & $3$ \\
        $f_{c}$ & Carrier frequency & $28$ [MHz] \\
        $B$ &  Bandwidth   &  $100$ [MHz] \\
        $N_0$ & Noise power density & $-173$ [dBm/Hz] \\ 
        $P_{\text{T}}$  &  Transmit power & $0:5:20$ [dBm] \\
        $d_{0}$ & Reference distance & $1$ [m] \\
       \bottomrule
    \end{tabular}
\end{table}

We evaluate the performance considering different numbers of active elements at RIS. Table~\ref{tab1:Setups} summarizes the parameters setup for the hybrid RIS architecture. 
We denote Setups 1 and 2 as the \emph{small} RIS, where $N_{\text{R}} = 32$ for both cases, but which differ from each other in the number of active elements, with $M \in \{2, 4\}$. Setups 3 and 4 are denoted as the \emph{large} RIS, in which we increase the number of elements at the RIS to $N_{\text{R}} = 64$, differing from each other in $M\in \{6, 8\}$. Also, due to the higher number of elements at the large RIS, we further increase the training overhead compared to the small RIS. We also consider $N_{\text{RF,R}}=M$ and $N_{\text{C,B}} = N_{\text{RF,B}}$ for all the hybrid setups in order to reduce the training overhead.

\begin{table} [h]
    \centering
    \caption{Parameters of the hybrid RIS.}
    \label{tab1:Setups}
    \begin{tabular}{lccccccccc}
    \toprule
    \textbf  & {$N_{\text{R}}$}& {$M$} & {$N_{\text{RF,R}}$} & {$K$}  & {$T$} & {$N_{\text{C,B}}$} &  {$N_{\text{RF,B}}$} &{$T_{\text{H}}$} \\
   \midrule
        {Setup 1}  & 32 &$4$ & $4$ & $5$ & $8$ & $8$ & $8$ & $40$\\
        {Setup 2}  & 32 &$2$ & $2$ & $5$ & $8$ & $8$ & $8$ & $40$ \\
        {Setup 3}  & 64 &$8$ & $8$ & $7$ & $8$ & $8$ & $8$ & $56$\\
        {Setup 4}  & 64 &$6$ & $6$ & $7$ & $8$ & $8$ & $8$ & $56$\\
   \bottomrule
    \end{tabular}
\end{table}

In addition, since we consider ANM in our proposed CE method, we assume the spatial frequencies are separated at least by $\left(\frac{4}{N_{\text{B}}}\right)$,  $\left(\frac{4}{N_{\text{R}}}\right)$, $\left(\frac{4}{N_{\text{M}}}\right)$ as to guarantee the super-resolution estimation~\cite{harnessing}. 

\subsection{Performance Metrics}
We evaluate the performance of MSEs~\footnote{The MSE can be also formulated by considering the sine of the angles/spatial frequencies. We clarify that the formulations are equivalent.} of the estimates of AoDs, AoAs, angle differences and products of propagation path gains, which are respectively given as
\begin{equation}
    \text{MSE}(\boldsymbol{\phi}_{\text{M,R}}) = \mathbb{E} \Bigg[ \frac{\| \boldsymbol{\phi}_{\text{M,R}} - \hat{\boldsymbol{\phi}}_{\text{M,R}}
    \|^{2}_{2}}{L_{\text{M,R}}} \Bigg],
\end{equation}
\begin{equation}
    \text{MSE}(\boldsymbol{\theta}_{\text{M,R}}) = \mathbb{E} \Bigg[ \frac{\| \boldsymbol{\theta}_{\text{M,R}} - \hat{\boldsymbol{\theta}}_{\text{M,R}}
    \|^{2}_{2}}{L_{\text{M,R}}} \Bigg],
\end{equation}
\begin{equation}
    \text{MSE}(\boldsymbol{\rho}_{\text{M,R}}) = \mathbb{E} \Bigg[ \frac{\| \boldsymbol{\rho}_{\text{M,R}} - \hat{\boldsymbol{\rho}}_{\text{M,R}}
    \|^{2}_{2}}{L_{\text{M,R}}} \Bigg],
\end{equation}
\begin{equation}
    \text{MSE}(\boldsymbol{\delta}) = \mathbb{E} \Bigg[ \frac{\|\boldsymbol{\delta} - \hat{\boldsymbol{\delta}}\|^{2}_2}{L_{\text{R,B}}L_{\text{M,R}}} \Bigg].
\end{equation}

The MSEs of $\boldsymbol{\phi}_{\text{R,B}}$, $\boldsymbol{\theta}_{\text{R,B}}$, $\boldsymbol{\rho}_{\text{R,B}}$ are defined in the same manner. We also evaluate the performance in terms of average effective SE in (bits/s/Hz) defined as~\cite{ngo2013energy}
\begin{equation}
    R = \mathbb{E} \Bigg[\frac{T_\text{c}-T_\text{H}}{T_\text{c}}  \log_{2} \Bigg(1 + \frac{ |\mathbf{w}^\mathsf{H}\hat{\bH}\mathbf{f}|^{2}}{\sigma^{2} + \text{var}(\mathbf{w}^\mathsf{H}\bH_{\text{e}}\mathbf{f})} \Bigg)\Bigg],
\end{equation} 
where $T_\text{c}$ is the number of time slots in a coherence time interval and $\bH_{\text{e}}$ denotes the channel estimation error, i.e., $\bH_{\text{e}} = \bH(\boldsymbol{\omega}^{*}) - \hat{\bH}$. We assume that the coherence time has 500 channel uses, i.e, $T_\text{c}$ = 500. 

\subsection{Comparison with the Passive RIS}
\begin{figure} [!t]
    \centering
    \includegraphics[width=0.6\columnwidth]{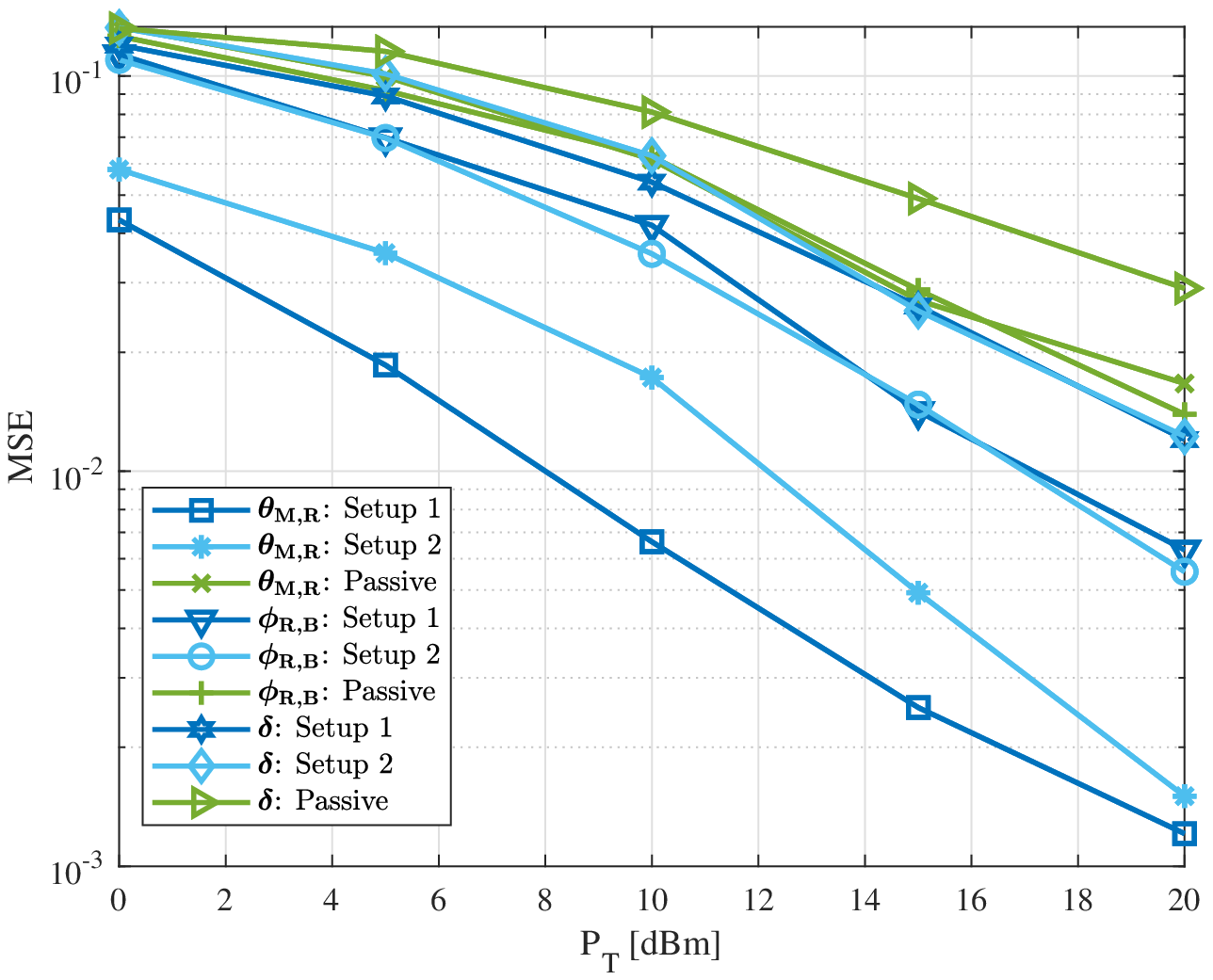}
    \caption{MSEs of the angular parameters for the RIS with $N_{\text{R}} = 32$ elements.}
    \label{fig:mse_1}
\end{figure}
\begin{figure} [!t]
    \centering
    \includegraphics[width=0.6\columnwidth]{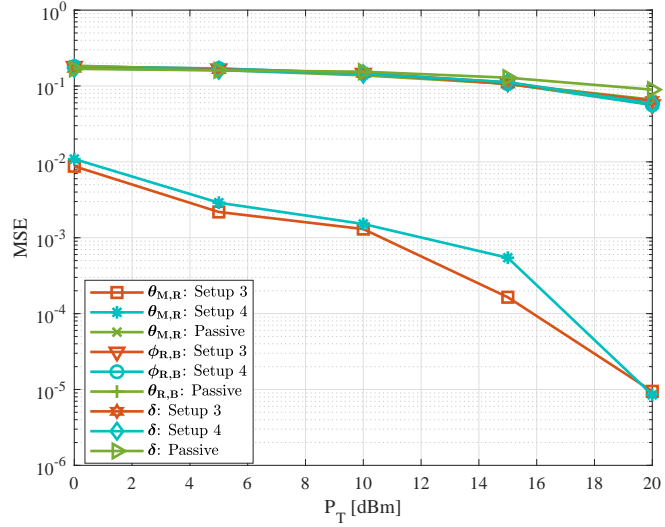}
    \caption{MSEs of the angular parameters for the RIS with $N_{\text{R}} = 64$ elements.}
    \label{fig:mse_2}
\end{figure}
The simulation results for the MSEs of channel parameters $\boldsymbol{\theta}_{\text{M,R}}$, $\boldsymbol{\phi}_{\text{R,B}}$ and $\boldsymbol{\delta}$ are shown in Fig.~\ref{fig:mse_1} for the small RIS. We consider $d_{\text{T}}=22$ [m], $d_{\text{x}}= 15$ [m], $d_{\text{y}}=2$ [m] and compare our results with the benchmark scheme, i.e., the passive RIS detailed in~\cite{he2020anm}. For the estimation of $\boldsymbol{\theta}_{\text{M,R}}$, all the setups of the hybrid RIS outperform the passive RIS. Setup 1 brings better performance than Setup 2, while Setup 2 has relatively lower power consumption due to the reduced number of RF chains at the RIS. The simulation results can be well explained by taking into account the path loss effect on the CE. In our proposed CE method, we perform CE at the RIS, where the path loss is proportional to the distance $d_{1}$.
\begin{figure} [!t]
    \centering
    \includegraphics[width=0.6\columnwidth]{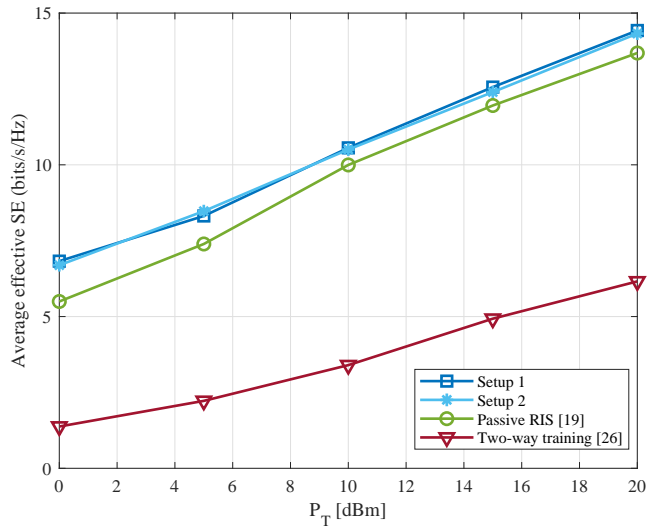}
    \caption{Overall analyses in terms of SE of our proposed method in comparison with the passive RIS~\cite{he2020anm} and hybrid RIS via two-way \textit{uplink downlink} training~\cite{schroeder2020passive}.}
    \label{fig:SE}
\end{figure}
However, when the CE is performed at the BS, the path loss is proportional to $d_{1}d_{2}$. That's why the proposed hybrid RIS CE can in general outperform the passive RIS CE.  
Regarding the MSE of $\boldsymbol{\delta}$, the hybrid RIS also has better performance than the passive RIS. For this parameter $\boldsymbol{\delta}$, the performance depends on the estimates from both the first and second stage. At the first stage, we can obtain better estimation of the channel parameters due to the lower path loss. However, at the second stage, the path loss is larger, which affects the estimation of $\boldsymbol{\rho}_{\text{R,B}}$. The overall performance heavily relies on the worse estimation, so the two setups donot have big performance gap, as shown in Fig.~\ref{fig:mse_1}.

We further evaluate the MSEs of the angular parameters for the large RIS with $64$ elements, i.e, Setups 3 and 4, in Fig.~\ref{fig:mse_2}. We consider $d_{\text{T}}=50$ [m], $d_{\text{x}}= 10$ [m], $d_{\text{y}}=2$ [m]. Since we consider a larger distance in the LoS link, the path loss of the MS-RIS-BS link brings more impact on the CE performance. Herein, we can clearly see that the estimate of $\boldsymbol{\theta}_{\text{M,R}}$ is better for the hybrid setups in~Fig.~\ref{fig:mse_2}. The MSEs of $\boldsymbol{\phi}_{\text{R,B}}$ and $\boldsymbol{\delta}$ show similar results as the small RIS in Fig.~\ref{fig:mse_1}. That is, the performance of both architectures is strongly affected by the path loss. 

\subsection{Overall Analyses of the Hybrid RIS}
We evaluate the overall performance in terms of SE of the Setups 1 and 2, and passive RIS via ANM~\cite{he2020anm} and that of the hybrid RIS via two-way \textit{uplink and downlink} training~\cite{schroeder2020passive}. We assume $M = N_\text{RF,R} = 10$ for the two-way training. For simplicity, we focus on the small RIS with $N_{\text{R}}= 32$ elements. The results of the average effective SE are provided in the Fig.~\ref{fig:SE}. From the figure, we observe that results of the hybrid setups and the passive RIS are aligned with the MSEs in Fig.~\ref{fig:mse_1}. Moreover, we observe that the overall performance of the Setups 1 and 2 are very similar, which means that we can use a reduced number of active elements and still achieve similar SE. The two-stage CE for hybrid RIS offers the possibility to improve the channel estimation of one of the individual channels, while the other channel still suffers from the effect of the path loss at the receiver. Thanks to the combination of efficient training and availability of measurements at the RIS, the overall performance of our proposed method is better than the passive RIS and the hybrid RIS via two-way uplink downlink training.

\subsubsection{CRLB}
\begin{figure} [!t]
    \centering
    \includegraphics[width=0.6\columnwidth]{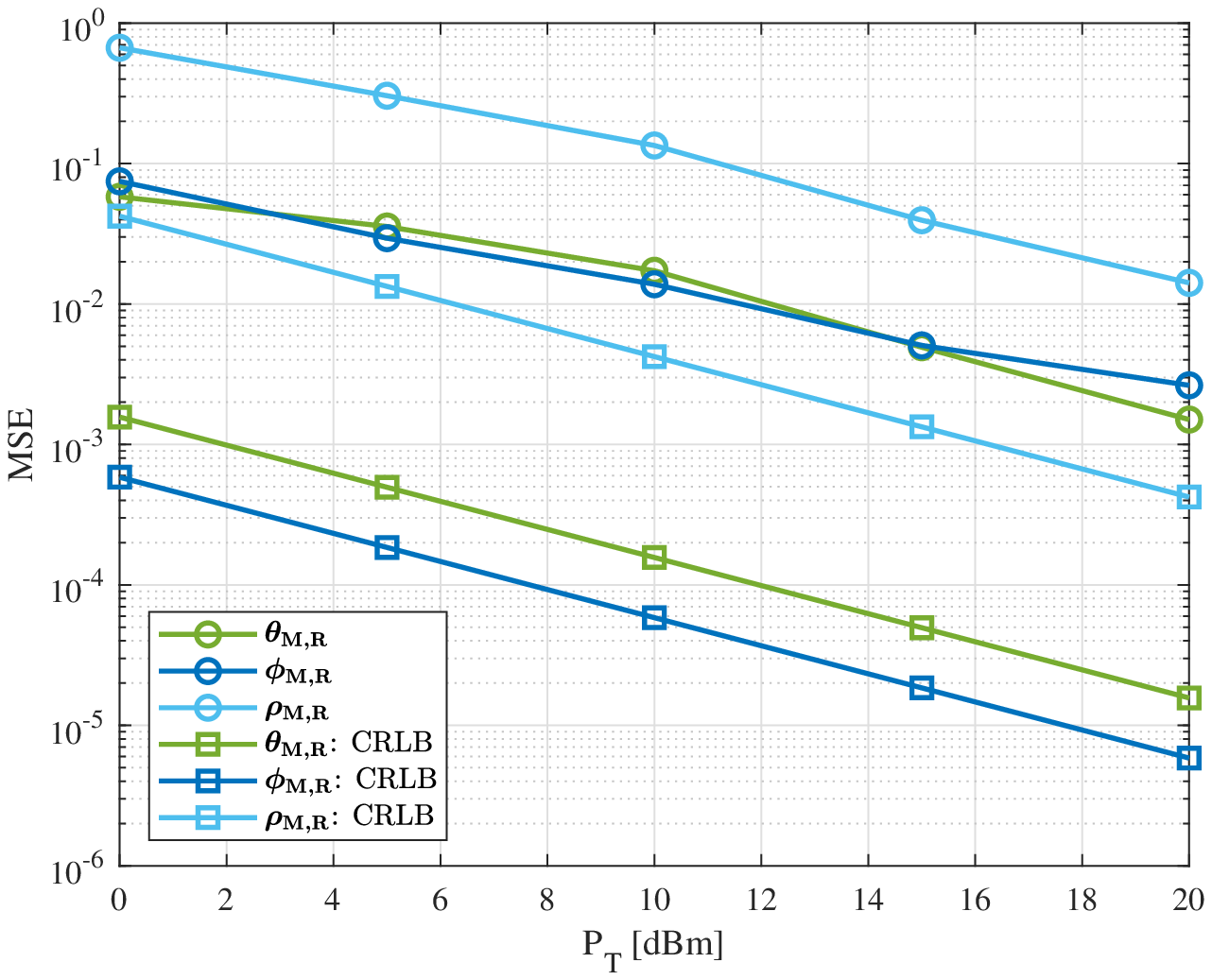}
    \caption{MSEs of the channel parameters in $\bH_\text{M,R}$ vs. CRLBs.}
    \label{fig:mse_5}
\end{figure}

\begin{figure} 
    \centering
    \includegraphics[width=0.6\columnwidth]{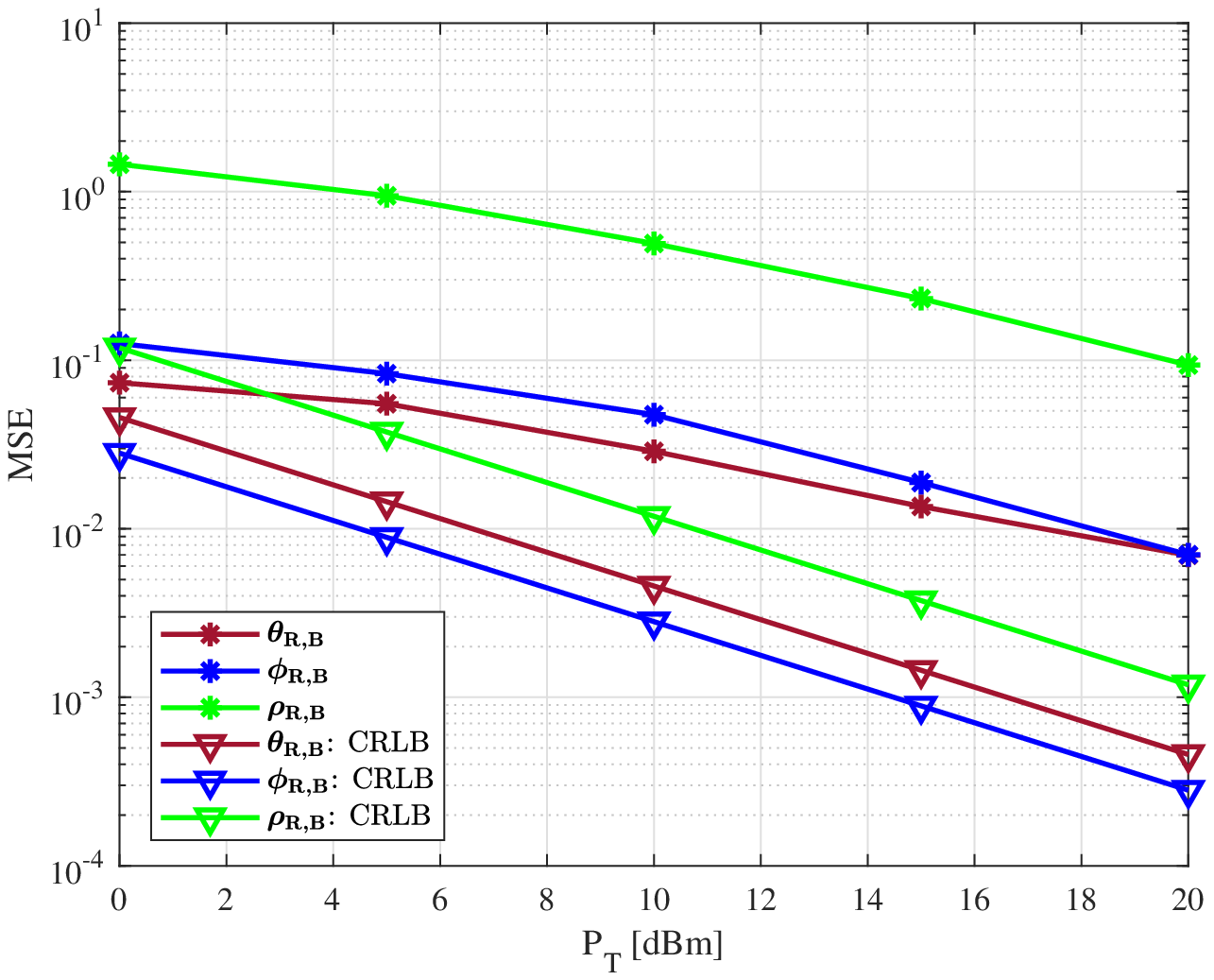}
    \caption{MSEs of the channel parameters in $\bH_\text{R,B}$ vs. CRLBs.}
    \label{fig:mse_6}
\end{figure}

We now examine the performance of our proposed method and the CRLB  (More details can be found in Section~\ref{section: crlb} with calculations/derivations drawn in the Appendixes~\ref{appendix:CRLB1} and~\ref{appendix:CRLB2}). Fig.~\ref{fig:mse_5} shows the MSEs of the channel parameters of the Setup 2, where $d_{\text{T}}=22$ [m], $d_{\text{x}}= 15$ [m], $d_{\text{y}}=2$ [m]. Despite the better performance obtained in comparison with the literature as show in Fig~\ref{fig:SE}, the gap between the MSEs of the proposed estimator and the CRLB shows that our method can still be further improved. Nevertheless, this improvement would inevitably come at the expense of more complexity, energy or time consumption, which is not desirable in this context.
In Fig.~\ref{fig:mse_6}, we evaluate the MSEs of the channel parameters $\boldsymbol{\theta}_{\text{R,B}}$, $\boldsymbol{\phi}_{\text{R,B}}$, $\boldsymbol{\rho}_{\text{R,B}}$ and make comparisons to their CRLBs. Similar to the previous case, we can see that the MSEs of the angles has a clear gap to the CRLBs, which may come from the assumption of perfect recovery of $\bH_\text{M,R}$ in the CRLB analyses. However, the MSE of the propagation path gain has a reduced gap to the lower bound. Comparing the results in Figs.~\ref{fig:mse_5} and~\ref{fig:mse_6}, we can see that the MSEs of the channel parameters are better at the first stage. This can be understood due to the reduced path loss at the hybrid RIS. Moreover, the estimation of the channel parameters $\boldsymbol{\theta}_{\text{R,B}}$, $\boldsymbol{\phi}_{\text{R,B}}$, $\boldsymbol{\rho}_{\text{R,B}}$ depends on the estimates of the previous stage. 

Again, note that we develop the CRLB for the second stage of CE by adopting  $\hat{\bH}_{\text{M,R}} \approx \bH_{\text{M,R}}$. In practice, the estimates of $\hat{\bH}_{\text{M,R}}$ depends on the training overhead, the number of RF chains at RIS, and the level of SNR. As a consequence, $\hat{\bH}_{\text{M,R}}$ may not be approximate to $\bH_{\text{M,R}}$. However, to characterize the channel (parameter) estimation errors at the first stage may yield a non-closed form solution of the CRLB. For this reason, we also leave this as future investigation.

\subsubsection{Location of the RIS}
\begin{figure} [!t]
    \centering
    \includegraphics[width=0.6\columnwidth]{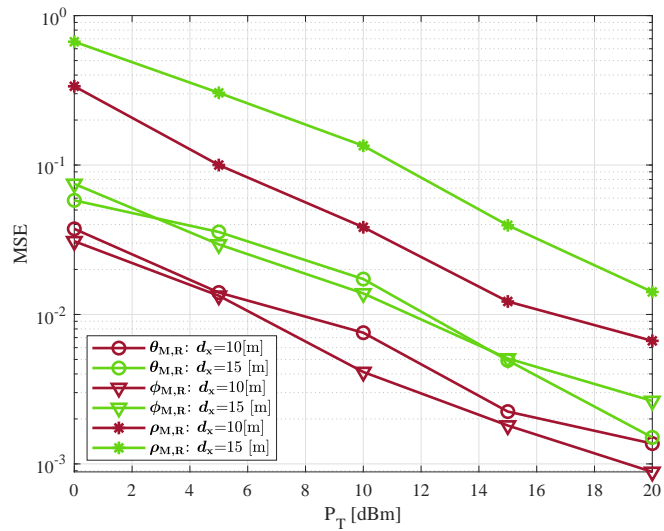}
    \caption{The effect of the location of the RIS for the channel parameter estimation of the Setup 2.}
    \label{fig:mse_3}
\end{figure}

\begin{figure} [!t]
    \centering
    \includegraphics[width=0.6\columnwidth]{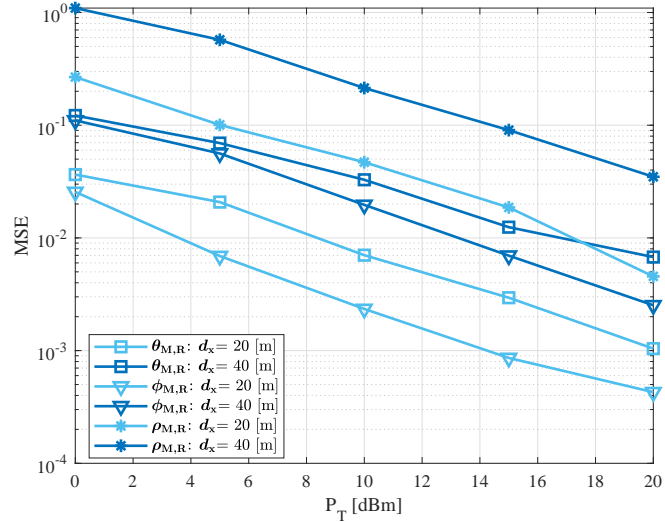}
    \caption{The effect of the location of the RIS for the channel parameter estimation considering the Setup 3.}
    \label{fig:mse_4}
\end{figure}

We also evaluate the effect of the position of the RIS on the CE of the first stage (i.e, recovery of $\bH_\text{M,R}$). Considering the Setup 2, the MSEs of the channel parameters $\boldsymbol{\theta}_{\text{M,R}}$, $\boldsymbol{\phi}_{\text{M,R}}$, $\boldsymbol{\rho}_{\text{M,R}}$ are shown in Fig.~\ref{fig:mse_3}. We consider $d_{\text{T}}=22$ [m], $d_{\text{x}}\in\{10, 15\}$ [m]. We can see that the estimation of the channel parameters for $d_{\text{x}}= 10$ [m] has better performance. 
We extend our study for the Setup 3, where the RIS has $N_{\text{R}}= 64$ elements. We evaluate the performance by considering $d_{\text{T}}=50$ [m], $d_{\text{y}}=2$ [m], and $d_{\text{x}}\in \{20, 40\}$ [m]. For the Setup 3, the distance between MS and RIS also affects the performance of the CE. We can see in Fig.~\ref{fig:mse_4} that the gap between the MSEs of the channel parameters for $d_{\text{x}}=20$ [m] and $d_{\text{x}}=40$ [m] is significant. 

\begin{figure} [h]
    \centering
    \includegraphics[width=0.6\columnwidth]{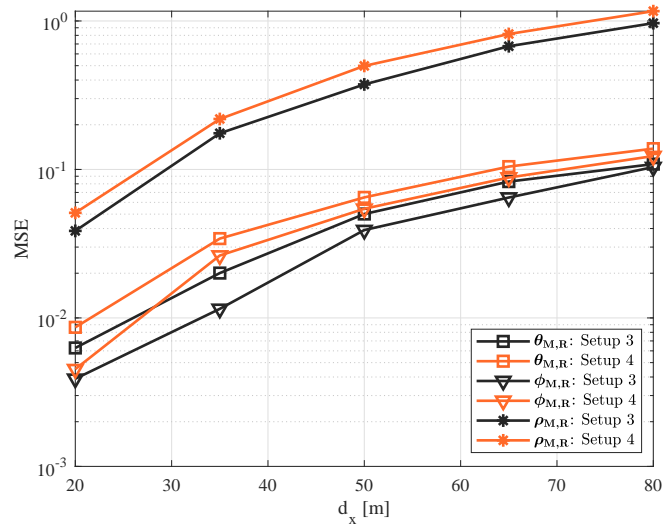}
    \caption{MSEs of the channel parameters for fixed $P_\text{T} = 10$ [dBm] and varying the distance RIS-MS ($d_{\text{x}}$).}
    \label{fig:mse_dist}
\end{figure}

Next, in Fig.~\ref{fig:mse_dist} we further evaluate the MSEs of the Setups 3 and 4 for a larger range of distance. We consider the transmission power fixed as $P_\text{T} = 10$ [dBm], $d_{\text{T}}=100$ [m] and the distance $d_{\text{x}}=\{20:15:80\}$ [m]. We can observe that the MSEs of the channel parameters increase significantly when the distance between RIS-MS~($d_{\text{x}}$) also increases. Moreover, the performance of the Setups 3 and 4 are quite similar to each other, which is aligned with the findings in Figs.~\ref{fig:mse_1} and~\ref{fig:mse_2}.  

The location of the RIS and how it affects the MSEs of the channel parameters depends on the number of elements/antennas at the BS, RIS and MS and also the distances between them. The overall performance may not be significantly affected by the position of the RIS. However, we clarify that the optimal location of the RIS can bring benefits from the practical deployment perspective. 

\subsubsection{Discussions}
We assume that the RIS feeds back the channel parameters to the BS by using the error-free backhaul link. Since we consider structured channel, the BS can reconstruct the MS-RIS channel matrix, i.e, $\hat{\bH}_{\text{M,R}}$, based on the estimates of the channel parameters. An alternative way is to feed back the entire channel matrix obtained from the equation~\eqref{eq_anm_stage1}. However, this assumption increases the complexity and volume of the feedback due to the higher dimensions of the channel matrix than these of channel parameters. Besides, by using the structured channel, we can exploit the inherent sparsity of the channel and reduce significantly the training overhead.

Furthermore, in this work we assume that the active elements are used to collect signal observations at RIS. We find that their positions donot affect the accuracy of the proposed CE algorithm as long as the total number of active elements is fixed. Nevertheless, the performance of CE can be further improved by increasing the training overhead and the number of RF chains and active elements. 

\section{Conclusions and Future Works}
\label{section: conclusions}
In this paper, we have studied the channel estimation via hybrid RIS for mmWave MIMO systems and proposed a two-stage CE procedure via ANM. Our results have shown that our proposed CE method could bring better performance than the passive RIS with the help of a small number of active elements. We have developed the analytical expressions of the CRLB for our proposed method, which provide another baselines in addition to passive RIS CE. It has been verified that the proposed hybrid RIS CE can outperform passive RIS CE under the same assumption of training overhead. 
We have shown that the location of the RIS can also affect the performance of the proposed hybrid RIS CE. 

The location of the RIS can be further studied for even larger number of elements/antennas, which is left as our future work. Future research directions also include the investigation of the CE performance for multi-user MIMO systems. In addition, hybrid RIS CE where the RIS has non-ideal hardware limitations would be another interesting future work. The assumption on knowing the exact number of paths and their order information as known \textit{priori} information should be relaxed in order to make the proposed method more applicable to the practical systems.  

\appendices
\section{Calculation of the CRLB: Stage 1 of CE}
\label{appendix:CRLB1}
In the following, we describe the calculations of the CRLB for the proposed CE method. At the first stage of CE, we estimate the parameters in the MS-RIS channel, defined as $\boldsymbol{\zeta} = \{\boldsymbol{\theta}_{\text{M,R}},\boldsymbol{\phi}_{\text{M,R}},\boldsymbol{\rho}_{\text{M,R}} \}$. The MSEs of the channel parameters in $\boldsymbol{\zeta}$ are lower bounded by $\text{CRLB}(\boldsymbol{\zeta})$, which is formulated as 
\begin{equation}
    \text{CRLB}(\boldsymbol{\zeta})\ge \bJ(\boldsymbol{\zeta})^{-1},
\end{equation}
with $\bJ(\boldsymbol{\zeta})$ being the FIM. For instance, the observation vector $\mathbf{y}_{\text{H}} = \mathrm{vec}(\bY_{\text{H}})$ follows Gaussian distribution with $\mathcal{CN}(\boldsymbol{\mu}_{1},\sigma^2\mathbf{I}_{MKT})$, where the mean $\boldsymbol{\mu}_{1}$ is expressed by
\begin{equation}
        \label{eq_m}
        \boldsymbol{\mu}_{1}= \sum\limits_{l = 1}^{L_{\text{M,R}}}[\boldsymbol{\rho}_{\text{M,R}}]_l(\mathbf{X}^{\mathsf{T}}\otimes\mathbf{W}_{\text{H}}) \big[\boldsymbol{\alpha}^*([\boldsymbol{\theta}_{\text{M,R}}]_l )\otimes\boldsymbol {\alpha}([\boldsymbol{\phi}_{\text{M,R}}]_l )\big].
\end{equation}
With this assumption, we can express the $(l,m)$th entry of the FIM as
\begin{equation}
    [\bJ(\boldsymbol{\zeta})]_{l,m} = \frac{2}{\sigma^2}\Re \Bigg\{\Bigg( \frac{\partial \boldsymbol{\mu}_{1}}{\partial [\boldsymbol{\zeta}]_{l}} \Bigg)^{\mathsf{H}} \Bigg( \frac{\partial \boldsymbol{\mu}_{1}}{\partial  [\boldsymbol{\zeta}]_{m}}\Bigg)\Bigg\}.
\end{equation}
In the next step, we describe the derivatives of $\boldsymbol{\mu}_{1}$ with respect to the channel parameters in $\boldsymbol{\zeta}$ and further derive the entries of the FIM. 

\subsection{Partial Derivatives}
First, we develop the partial derivatives of $\boldsymbol{\mu}_{1}$ with respect to the channel parameter $\boldsymbol{\theta}_{\text{M,R}}$.   
\begin{equation}
    \label{eq:derivate_m_theta1}
    \begin{aligned}
            \Bigg(\frac{\partial \boldsymbol{\mu}_{1}}{\partial [\boldsymbol{\theta}_{\text{M,R}}]_l} \Bigg) = &(\mathbf{X}^{\mathsf{T}}\otimes\mathbf{W}_{\text{H}})([\boldsymbol{\rho}_{\text{M,R}}]_l) j \pi \cos([\boldsymbol{\theta}_{\text{M,R}}]_l) \\
            & \big[\tilde{\boldsymbol{\alpha}}^{*}([\boldsymbol{\theta}_{\text{M,R}}]_l)\otimes\boldsymbol{\alpha}([\boldsymbol{\phi}_{\text{M,R}}]_l)\big],
    \end{aligned}
\end{equation}
where $[\tilde{\boldsymbol{\alpha}}([\boldsymbol{\theta}_{\text{M,R}}]_l)]_{n}=(n-1)\exp\{j\pi(n-1)\sin( [\boldsymbol{\theta}_{\text{M,R}}]_l)\}$ for $n=1,\cdots,N_\text{M}$. Similarly, we obtain the partial derivatives with respect to the remaining parameters as follows 
\begin{equation}
    \begin{aligned}
        \Bigg( \frac{\partial \boldsymbol{\mu}_{1}}{\partial [\boldsymbol{\phi}_{\text{M,R}}]_l} \Bigg) = & (\mathbf{X}^{\mathsf{T}}\otimes\mathbf{W}_{\text{H}})([\boldsymbol{\rho}_{\text{M,R}}]_l) j \pi \cos([\boldsymbol{\phi}_{\text{M,R}}]_l) \\
        &\big[\boldsymbol{\alpha}^*([\boldsymbol{\theta}_{\text{M,R}}]_l)
        \otimes \tilde{\boldsymbol{\alpha}}([\boldsymbol{\phi}_{\text{M,R}}]_l)\big], 
    \end{aligned}
\end{equation}
\begin{equation}
        \begin{aligned}
            \Bigg( \frac{\partial \boldsymbol{\mu}_{1}}{\partial [\boldsymbol{\rho}_{\text{M,R}}]_l} \Bigg)= & (\mathbf{X}^{\mathsf{T}}\otimes\mathbf{W}_{\text{H}})\big[\boldsymbol{\alpha}^*([\boldsymbol{\theta}_{\text{M,R}}]_l)\otimes \boldsymbol{\alpha}([\boldsymbol{\phi}_{\text{M,R}}]_l)\big],
        \end{aligned}
\end{equation}
where $[\tilde{\boldsymbol{\alpha}}([\boldsymbol{\phi}_{\text{M,R}}]_l)]_{n}=(n-1)\exp\{j\pi(n-1)\sin( [\boldsymbol{\phi}_{\text{M,R}}]_l)\}$ for $n=1,\cdots,N_\text{R}$. The derivatives $\Big(\frac{\partial \boldsymbol{\mu}_{1}}{\partial [\boldsymbol{\theta}_{\text{M,R}}]_l} \Big)^{\mathsf{H}}$, $\Big(\frac{\partial \boldsymbol{\mu}_{1}}{\partial [\boldsymbol{\phi}_{\text{M,R}}]_l} \Big)^{\mathsf{H}}$, and $\Big(\frac{\partial \boldsymbol{\mu}_{1}}{\partial [\boldsymbol{\rho}_{\text{M,R}}]_l} \Big)^{\mathsf{H}}$ can be obtained by following the equations developed above and by applying the conjugate transpose. 

\subsection{Calculation of Fisher Information Matrix}
We first calculate the entries of the elements in the principal diagonal. For instance, the $(l,l)$th entry of $\bJ(\boldsymbol{\zeta})$ is given by
\begin{equation}
    \label{eq:prim_deriv}
         \bJ({[\boldsymbol{\theta}_{\text{M,R}}]_l,[\boldsymbol{\theta}_{\text{M,R}}]_l})  =  
         \frac{2}{\sigma^2}\Bigg\{\Bigg( \frac{\partial \boldsymbol{\mu}_{1}}{\partial [\boldsymbol{\theta}_{\text{M,R}}]_l} \Bigg)^{\mathsf{H}} \Bigg( \frac{\partial \boldsymbol{\mu}_{1}}{\partial  [\boldsymbol{\theta}_{\text{M,R}}]_l}\Bigg)\Bigg\}. 
\end{equation}
By substituting the derivatives of $\boldsymbol{\mu}_{1}$ with respect to $\boldsymbol{\theta}_{\text{M,R}}$, we obtain
\begin{equation}
    \bJ({[\boldsymbol{\theta}_{\text{M,R}}]_l,[\boldsymbol{\theta}_{\text{M,R}}]_l}) = \frac{2}{\sigma^2}\Bigg\{([\boldsymbol{\rho}_{\text{M,R}}]_l)^{2}\pi^{2}\cos^{2}([\boldsymbol{\theta}_{\text{M,R}}]_l)\big[\boldsymbol{\chi}_{1}\boldsymbol{\xi}_{1}\boldsymbol{\kappa}_{1}\big]\Bigg\},
\end{equation}
where $\boldsymbol{\chi}_{1} = \big[\tilde{\boldsymbol{\alpha}}^{\mathsf{T}}([\boldsymbol{\theta}_{\text{M,R}}]_l)\otimes\boldsymbol{\alpha}^{\mathsf{H}}([\boldsymbol{\phi}_{\text{M,R}}]_l)\big]$, $\boldsymbol{\xi}_{1} = (\mathbf{X}^{\mathsf{T}}\otimes\mathbf{W}_{\text{H}})^{\mathsf{H}}(\mathbf{X}^{\mathsf{T}}\otimes\mathbf{W}_{\text{H}})$, and $\boldsymbol{\kappa}_{1}=\big[
\tilde{\boldsymbol{\alpha}}^{*}([\boldsymbol{\theta}_{\text{M,R}}]_l)\otimes\boldsymbol{\alpha}([\boldsymbol{\phi}_{\text{M,R}}]_l)\big]$.
Similarly, we can obtain  
\begin{equation}
    \begin{aligned}
        \bJ({[\boldsymbol{\phi}_{\text{M,R}}]_l,[\boldsymbol{\phi}_{\text{M,R}}]_l}) & =
        \frac{2}{\sigma^2}\Bigg\{\Bigg( \frac{\partial \boldsymbol{\mu}_{1}}{\partial [\boldsymbol{\phi}_{\text{M,R}}]_l} \Bigg)^{\mathsf{H}} \Bigg( \frac{\partial \boldsymbol{\mu}_{1}}{\partial  [\boldsymbol{\phi}_{\text{M,R}}]_l}\Bigg)\Bigg\} \\
        & =\frac{2}{\sigma^2}([\boldsymbol{\rho}_{\text{M,R}}]_l)^{2}\pi^{2}\cos^{2}([\boldsymbol{\phi}_{\text{M,R}}]_l)\big[\boldsymbol{\chi}_{2}\boldsymbol{\xi}_{1}\boldsymbol{\kappa}_{2}\big],
    \end{aligned}
\end{equation} 
\begin{equation}
    \begin{aligned}
        \bJ({[\boldsymbol{\rho}_{\text{M,R}}]_l,[\boldsymbol{\rho}_{\text{M,R}}]_l}) & = 
        \frac{2}{\sigma^2}\Bigg\{\Bigg( \frac{\partial \boldsymbol{\mu}_{1}}{\partial [\boldsymbol{\rho}_{\text{M,R}}]_l} \Bigg)^{\mathsf{H}} \Bigg( \frac{\partial \boldsymbol{\mu}_{1}}{\partial  [\boldsymbol{\rho}_{\text{M,R}}]_l}\Bigg)\Bigg\} \\
        & = \frac{2}{\sigma^2}\big[\boldsymbol{\chi}_{3}\boldsymbol{\xi}_{1}\boldsymbol{\kappa}_{3}\big],
    \end{aligned}
\end{equation}
where $\boldsymbol{\chi}_{2}=\big[\boldsymbol{\alpha}^{\mathsf{T}}([\boldsymbol{\theta}_{\text{M,R}}]_l)\otimes \tilde{\boldsymbol{\alpha}}^{\mathsf{H}}([\boldsymbol{\phi}_{\text{M,R}}]_l)\big]$, 
$\boldsymbol{\kappa}_{2}= \big[\boldsymbol{\alpha}^*([\boldsymbol{\theta}_{\text{M,R}}]_l)\otimes \tilde{\boldsymbol{\alpha}}([\boldsymbol{\phi}_{\text{M,R}}]_l)\big]$, $\boldsymbol{\chi}_{3}=\big[\boldsymbol{\alpha}^{\mathsf{T}}([\boldsymbol{\theta}_{\text{M,R}}]_l)\otimes \boldsymbol{\alpha}^{\mathsf{H}}([\boldsymbol{\phi}_{\text{M,R}}]_l)\big]$, and $\boldsymbol{\kappa}_{3}=\big[\boldsymbol{\alpha}^*([\boldsymbol{\theta}_{\text{M,R}}]_l)\otimes \boldsymbol{\alpha}([\boldsymbol{\phi}_{\text{M,R}}]_l)\big]$.
Note that the calculations of the $(l,m)$th off-diagonal entries follow the same procedure.

\section{Calculation of the CRLB: Stage 2 of CE}
\label{appendix:CRLB2}
At the second CE stage, we target at recovering the parameters of the RIS-BS channel, defined as $\boldsymbol{\eta} = \{ \boldsymbol{\theta}_{\text{R,B}},\boldsymbol{\phi}_{\text{R,B}}, \boldsymbol{\rho}_{\text{R,B}}\}$. Note that the procedure to obtain the elements of the FIM and the corresponding CRLB is similar to the analysis in Appendix~\ref{appendix:CRLB1}. We assume that the observation vector $\mathbf{y} = \mathrm{vec}(\bY)$ follows Gaussian distribution with $\mathcal{CN}(\boldsymbol{\mu}_{2},\sigma^2\mathbf{I}_{T K N_{\text{C,B}}})$, where $\boldsymbol{\mu}_{2}$ is defined by 
\begin{equation}
      \boldsymbol{\mu}_{2} =\sum\limits_{l = 1}^{L_{\text{R,B}}}[\boldsymbol{\rho}_{\text{R,B}}]_l(\tilde{\bU}^{\mathsf{T}}\otimes \mathbf{W}_{\text{B}}^{\mathsf{H}}) \big(\boldsymbol{\alpha}^*([\boldsymbol{\theta}_{\text{R,B}}]_l )\otimes\boldsymbol {\alpha}([\boldsymbol{\phi}_{\text{R,B}}]_l )\big).
\end{equation}
By adopting this assumption, we can formulate the FIM $\bJ(\boldsymbol{\eta})$ as 
\begin{equation}
    [\bJ(\boldsymbol{\eta})]_{l_{1},l_{2}} = \frac{2}{\sigma^2}\Re \Bigg\{\Bigg( \frac{\partial \boldsymbol{\mu}_{2}}{\partial [\boldsymbol{\eta}]_{l_{1}}} \Bigg)^{\mathsf{H}} \Bigg( \frac{\partial \boldsymbol{\mu}_{2}}{\partial  [\boldsymbol{\eta}]_{l_{2}}}\Bigg)\Bigg\}.
\end{equation}
In the following, we develop the derivatives of $\boldsymbol{\mu}_{2}$ with respect to $\boldsymbol{\eta}$ in detail. 

\subsection{Partial Derivatives}
We describe the derivatives of $\boldsymbol{\mu}_{2}$ with respect to the channel parameters in $\boldsymbol{\eta}$. 
\begin{align}
    \label{deriv_2_channel_theta}
    \Bigg(\frac{\partial \boldsymbol{\mu}_{2}}{\partial [\boldsymbol{\theta}_{\text{R,B}}]_{l_{1}}}\Bigg) =  & (\tilde{\bU}^{\mathsf{T}}\otimes \mathbf{W}_{\text{B}}^{\mathsf{H}})([\boldsymbol{\rho}_{\text{R,B}}]_l) j \pi \cos([\boldsymbol{\theta}_{\text{R,B}}]_{l_{1}}) \\
    \big[
    &\tilde{\boldsymbol{\alpha}}^{*}([\boldsymbol{\theta}_{\text{R,B}}]_{l_{1}})\otimes\boldsymbol{\alpha}([\boldsymbol{\phi}_{\text{R,B}}]_{l_{1}})\big],
\end{align}
\begin{align}
     \label{deriv_2_channel_phi}
        \Bigg(\frac{\partial \boldsymbol{\mu}_{2}}{\partial [\boldsymbol{\phi}_{\text{R,B}}]_{l_{1}}}\Bigg) = & (\tilde{\bU}^{\mathsf{T}}\otimes \mathbf{W}_{\text{B}}^{\mathsf{H}})([\boldsymbol{\rho}_{\text{R,B}}]_{l_{1}}) j \pi \cos([\boldsymbol{\phi}_{\text{R,B}}]_{l_{1}}) \\ 
        &\big[\boldsymbol{\alpha}^*([\boldsymbol{\theta}_{\text{R,B}}]_{l_{1}})
        \otimes \tilde{\boldsymbol{\alpha}}([\boldsymbol{\phi}_{\text{R,B}}]_{l_{1}})\big],
\end{align}
\begin{align}
         \label{deriv_2_channel_rho}
        \Bigg(\frac{\partial \boldsymbol{\mu}_{2}}{\partial [\boldsymbol{\rho}_{\text{R,B}}]_{l_{1}}} \Bigg) = & (\tilde{\bU}^{\mathsf{T}}\otimes \mathbf{W}_{\text{B}}^{\mathsf{H}}) \big[\boldsymbol{\alpha}^*([\boldsymbol{\theta}_{\text{R,B}}]_{l_{1}})\otimes \boldsymbol{\alpha}([\boldsymbol{\phi}_{\text{R,B}}]_{l_{1}})\big],
\end{align}
where $[\tilde{\boldsymbol{\alpha}}([\boldsymbol{\phi}_{\text{R,B}}]_{l_{1}})]_{n}=(n-1)\exp\{j\pi(n-1)\sin( [\boldsymbol{\phi}_{\text{R,B}}]_{l_{1}})\}$ for $n=1,\cdots,N_\text{B}$ and $[\tilde{\boldsymbol{\alpha}}([\boldsymbol{\theta}_{\text{R,B}}]_{l_{1}})]_{n}=(n-1)\exp\{j\pi(n-1)\sin( [\boldsymbol{\theta}_{\text{R,B}}]_{l_{1}})\}$ for $n=1,\cdots,N_\text{R}$.

\subsection{Calculation of Fisher Information Matrix}
For sake of brevity, we only define the entries of the elements on the main diagonal of the FIM as
\begin{align}
    \label{J_theta_2channel}
    \bJ({[\boldsymbol{\theta}_{\text{R,B}}]_{l_{1}},[\boldsymbol{\theta}_{\text{R,B}}]_{l_{1}}}) = \frac{2}{\sigma^2}\Bigg\{\Bigg( \frac{\partial \boldsymbol{\mu}_{2}}{\partial [\boldsymbol{\theta}_{\text{R,B}}]_{l_{1}}} \Bigg)^{\mathsf{H}} \Bigg( \frac{\partial \boldsymbol{\mu}_{2}}{\partial  [\boldsymbol{\theta}_{\text{R,B}}]_{l_{1}}}\Bigg)\Bigg\}, 
\end{align}
\begin{align}
        \label{J_phi_2channel}
    \bJ({[\boldsymbol{\phi}_{\text{R,B}}]_{l_{1}},[\boldsymbol{\phi}_{\text{R,B}}]_{l_{1}}}) =  \frac{2}{\sigma^2}\Bigg\{\Bigg( \frac{\partial \boldsymbol{\mu}_{2}}{\partial [\boldsymbol{\phi}_{\text{R,B}}]_{l_{1}}} \Bigg)^{\mathsf{H}} \Bigg( \frac{\partial \boldsymbol{\mu}_{2}}{\partial  [\boldsymbol{\phi}_{\text{R,B}}]_{l_{1}}}\Bigg)\Bigg\}, 
\end{align}
\begin{align}
    \label{J_rho_2channel}
    \bJ({[\boldsymbol{\rho}_{\text{R,B}}]_{l_{1}},[\boldsymbol{\rho}_{\text{R,B}}]_{l_{1}}}) = \frac{2}{\sigma^2}\Bigg\{\Bigg( \frac{\partial \boldsymbol{\mu}_{2}}{\partial [\boldsymbol{\theta}_{\text{R,B}}]_{l_{1}}} \Bigg)^{\mathsf{H}} \Bigg( \frac{\partial \boldsymbol{\mu}_{2}}{\partial  [\boldsymbol{\theta}_{\text{R,B}}]_{l_{1}}}\Bigg)\Bigg\}. 
\end{align}
By applying the derivatives of $\boldsymbol{\mu}_{2}$ with respect to $\boldsymbol{\eta}$, we obtain
\begin{equation}
    \begin{split}
    \bJ({[\boldsymbol{\theta}_{\text{R,B}}]_{l_{1}},[\boldsymbol{\theta}_{\text{R,B}}]_{l_{1}}}) = &  \frac{2}{\sigma^2}\Bigg\{([\boldsymbol{\rho}_{\text{R,B}}]_{l_{1}})^{2}\pi^{2}\cos^{2}([\boldsymbol{\theta}_{\text{R,B}}]_{l_{1}}) \big[\boldsymbol{\chi}_{4}\boldsymbol{\xi}_{2}\boldsymbol{\kappa}_{4}\big]\Bigg\},
    \end{split}
\end{equation}
where $\boldsymbol{\chi}_{4}=\big[\boldsymbol{\alpha}^{\mathsf{T}}([\boldsymbol{\theta}_{\text{R,B}}]_{l_{1}})\otimes \tilde{\boldsymbol{\alpha}}^{\mathsf{H}}([\boldsymbol{\phi}_{\text{R,B}}]_{l_{1}})\big]$, $\boldsymbol{\xi}_{2} =(\tilde{\bU}^{\mathsf{T}}\otimes \mathbf{W}_{\text{B}}^{\mathsf{H}})^{\mathsf{H}}(\tilde{\bU}^{\mathsf{T}}\otimes \mathbf{W}_{\text{B}}^{\mathsf{H}})$, and $\boldsymbol{\kappa}_{4}= \big[\boldsymbol{\alpha}^*([\boldsymbol{\theta}_{\text{R,B}}]_{l_{1}})\otimes \tilde{\boldsymbol{\alpha}}([\boldsymbol{\phi}_{\text{R,B}}]_{l_{1}})\big]$. In a similar procedure, we can re-write~\eqref{J_phi_2channel} and~\eqref{J_rho_2channel}, respectively, as
\begin{equation}
    \bJ({[\boldsymbol{\theta}_{\text{R,B}}]_{l_{1}},[\boldsymbol{\theta}_{\text{R,B}}]_{l_{1}}}) = \frac{2}{\sigma^2}([\boldsymbol{\rho}_{\text{R,B}}]_{l_{1}})^{2}\pi^{2}\cos^{2}([\boldsymbol{\phi}_{\text{R,B}}]_{l_{1}})\big[\boldsymbol{\chi}_{5}\boldsymbol{\xi}_{2}\boldsymbol{\kappa}_{5}\big],
\end{equation}
\begin{equation}
    \bJ({[\boldsymbol{\rho}_{\text{R,B}}]_{l_{1}},[\boldsymbol{\rho}_{\text{R,B}}]_{l_{1}}}) = \frac{2}{\sigma^2}\big[\boldsymbol{\chi}_{6}\boldsymbol{\xi}_{2}\boldsymbol{\kappa}_{6}\big],
\end{equation}
where $\boldsymbol{\chi}_{5} = \big[\boldsymbol{\alpha}^{\mathsf{T}}([\boldsymbol{\theta}_{\text{R,B}}]_{l_{1}})\otimes \tilde{\boldsymbol{\alpha}}^{\mathsf{H}}([\boldsymbol{\phi}_{\text{R,B}}]_{l_{1}})\big]$,
$\boldsymbol{\kappa}_{5}= \big[\boldsymbol{\alpha}^*([\boldsymbol{\theta}_{\text{R,B}}]_{l_{1}})\otimes \tilde{\boldsymbol{\alpha}}([\boldsymbol{\phi}_{\text{R,B}}]_{l_{1}})\big]$, $\boldsymbol{\chi}_{6}=\big[\boldsymbol{\alpha}^{\mathsf{T}}([\boldsymbol{\theta}_{\text{R,B}}]_{l_{1}})\otimes \boldsymbol{\alpha}^{\mathsf{H}}([\boldsymbol{\phi}_{\text{R,B}}]_{l_{1}})\big]$, and $\boldsymbol{\kappa}_{6}=\big[\boldsymbol{\alpha}^*([\boldsymbol{\theta}_{\text{R,B}}]_{l_{1}})\otimes \boldsymbol{\alpha}([\boldsymbol{\phi}_{\text{R,B}}]_{l_{1}})\big]$. The remaining entries of the FIM can be derived by following the same procedure.  

\bibliographystyle{IEEEtran}
\bibliography{IEEEabrv,references3}
\end{document}